\documentclass[numberedappendix]{emulateapj}
\usepackage{graphicx}
\usepackage{amsmath}

\usepackage[toc,page]{appendix}

\usepackage{color}

\newcommand{\appropto}{\mathrel{\vcenter{
  \offinterlineskip\halign{\hfil$##$\cr
    \propto\cr\noalign{\kern2pt}\sim\cr\noalign{\kern-2pt}}}}}

\def\kms{\ifmmode{\rm km\thinspace s^{-1}}\else km\thinspace s$^{-1}$\fi}

\shortauthors{Rappaport et al.~2013}
\shorttitle{KOI-2700 -- dusty-tail planet}

\begin{document}

% ------------------------------------------------------------------------
% New commands
%
\def\ltsima{$\; \buildrel < \over \sim \;$}
\def\lsim{\lower.5ex\hbox{\ltsima}}
\def\gtsima{$\; \buildrel > \over \sim \;$}
\def\gsim{\lower.5ex\hbox{\gtsima}}
% -------------------------------------------------------------------------
%

\bibliographystyle{apj}

\title{
KOI-2700b -- a Planet Candidate with Dusty Effluents on a 22-Hour Orbit
}

\author{
Saul~Rappaport\altaffilmark{1},
Thomas Barclay\altaffilmark{2},
John DeVore\altaffilmark{3},
Jason Rowe\altaffilmark{4},
Roberto~Sanchis-Ojeda\altaffilmark{1},
Martin Still\altaffilmark{2}
}

\altaffiltext{1}{Department of Physics, and Kavli Institute for
  Astrophysics and Space Research, Massachusetts Institute of
  Technology, Cambridge, MA 02139, USA; sar@mit.edu, rsanchis86@gmail.com}

\altaffiltext{2}{BAER Institute/NASA Ames Research Center, M/S 244-30, Moffett Field, Mountain View, California 94035; thomas.barclay@nasa.gov, martin.d.still@nasa.gov}

\altaffiltext{3}{Visidyne, Inc., 111 South Bedford St., Suite \#103, Burlington, MA 01803; devore@visidyne.com}

\altaffiltext{4}{SETI Institute,189 Bernardo Ave, Suite 100
Mountain View, CA 94043; NASA Ames Research Center, M/S 244-30, Moffett Field, Mountain View, California 94035;  jasonfrowe@gmail.com}

\begin{abstract}

  {\em Kepler} planet candidate KOI-2700b (KIC 8639908b) with an orbital period of 21.84 hours exhibits a distinctly asymmetric transit profile, likely indicative of the emission of dusty effluents, and reminiscent of KIC 1255b.  The host star has $T_{\rm eff} = 4435 $ K, $M \simeq 0.63 \, M_\odot$, and $R \simeq 0.57 \, R_\odot$, comparable to the parameters ascribed to KIC 12557548.  The transit egress can be followed for $\sim$25\% of the orbital period, and, if interpreted as extinction from a dusty comet-like tail, indicates a long lifetime for the dust grains of more than a day.  We present a semi-physical model for the dust tail attenuation, and fit for the physical parameters contained in that expression.  The transit is not sufficiently deep to allow for a study of the transit-to-transit variations, as is the case for KIC 1255b; however, it is clear that the transit depth is slowly monotonically decreasing by a factor of $\sim$2 over the duration of the {\em Kepler} mission.  The existence of a second star hosting a planet with a dusty comet-like tail would help to show that such objects may be more common and less exotic than originally thought.  According to current models, only quite small planets with $M_p \lesssim 0.03\,M_\oplus$ are likely to release a detectable quantity of dust.  Thus, any ``normal-looking'' transit that is inferred to arise from a rocky planet of radius greater than $\sim$$1/2\,R_\oplus$ should not exhibit any hint of a dusty tail.  Conversely, if one detects an asymmetric transit, due to a dusty tail, then it will be very difficult to detect the hard body of the planet within the transit because, by necessity, the planet must be quite small (i.e., $\lesssim 0.3 \,R_\oplus$).      

\end{abstract}

\keywords{planetary systems---planets and satellites: detection---planets and satellites: individual (KOI-2700b)}

\section{Introduction}

Over the past four years some 3600 planet candidates have been found with the {\em Kepler} mission (e.g., Borucki et al.~2011; Batalha et al.~2013; Burke et al.~2013).  Approximately 1350 of these reside in multiplanet systems.  The observed orbital period distribution peaks at $\sim$4 days and 80\% of the planets are found in the orbital period range $P_{\rm orb} = 2.5-70$ days.  The falloff toward longer periods is due to observational selection effects, i.e., the probability of a transit decreases roughly as $P_{\rm orb}^{-2/3}$, and the duty cycle of the transits has the same dependence on $P_{\rm orb}$.  Toward the short-period end, planets of an Earth radius or larger have a {\em higher} detection probability because of the increasing number of transits observed, and yet their numbers are falling due to non-observational selection effects.  

\begin{figure*}
\begin{center}
\includegraphics[width=0.47 \textwidth]{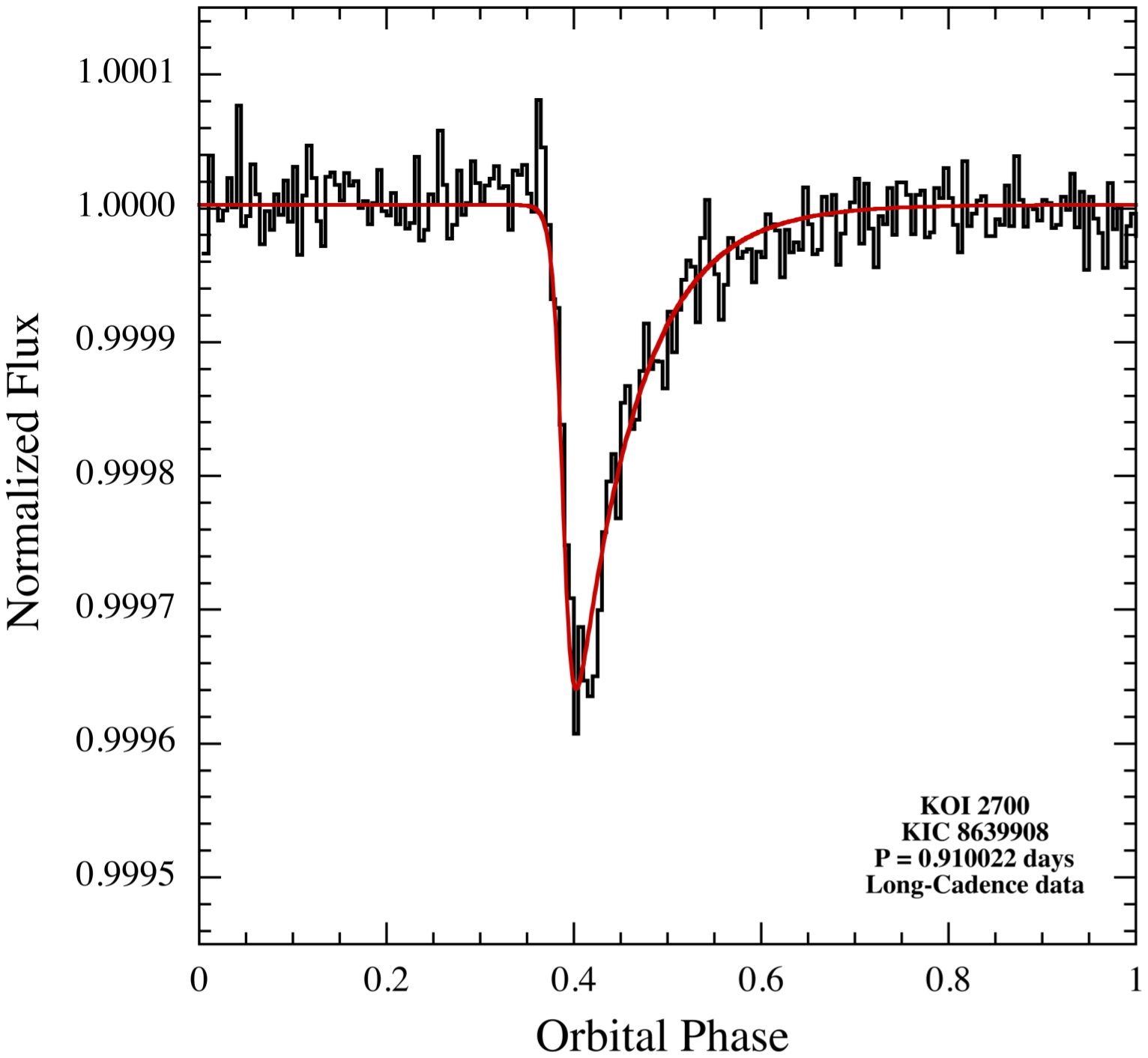} \hglue0.3cm
\includegraphics[width=0.475 \textwidth]{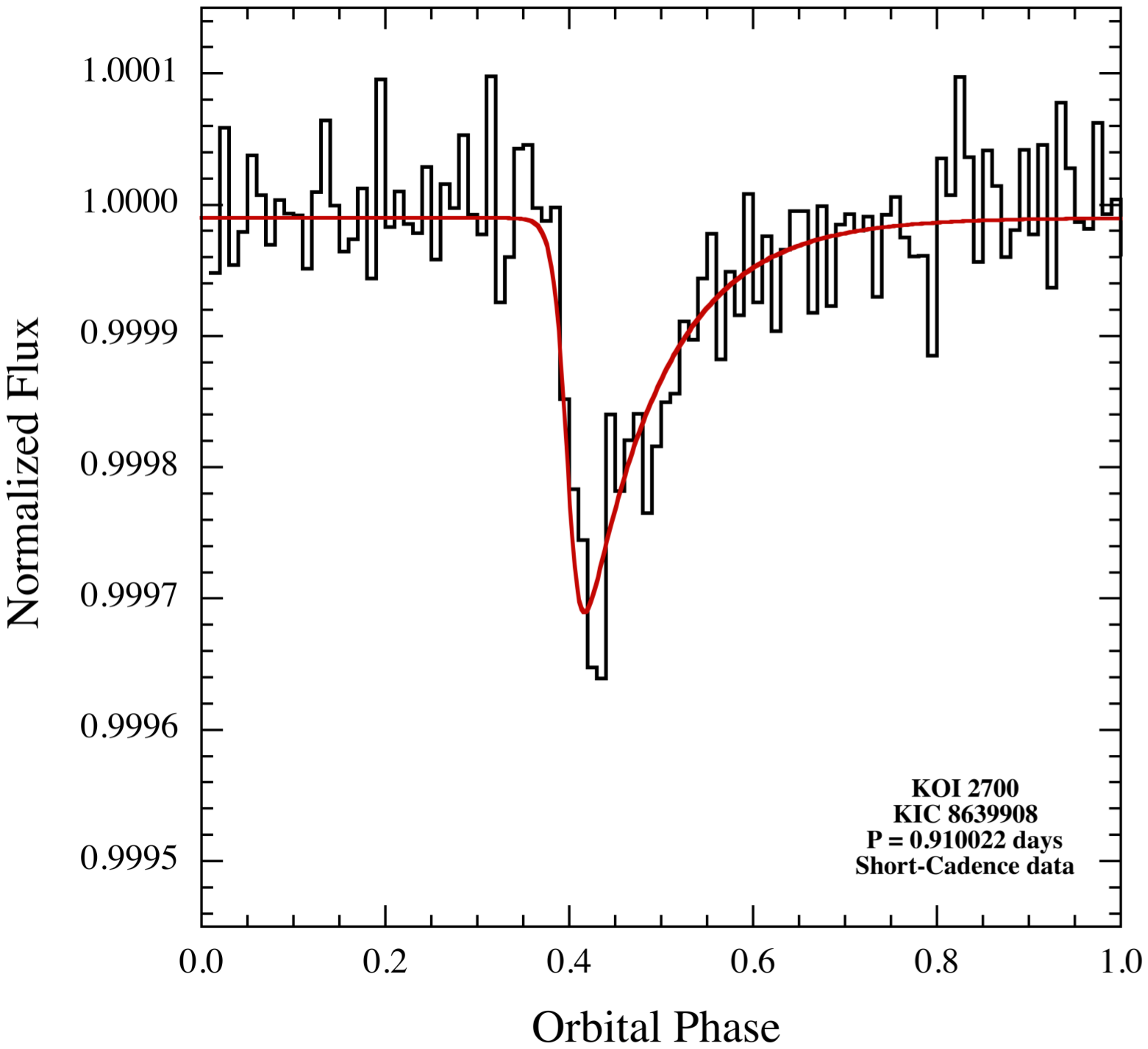}
\caption{{\it Left}.---Folded light curve of KOI-2700b for 16 quarters of long-cadence data in 200 bins (6.5-min.~each).  The red curve is a fit to a non-physical model (see text).  
{\it Right.}---Same as left panel, but using 2 quarters of short-cadence data in 100 bins (13-min.~each).  The red curve is a fit to the same non-physical model as in the left panel. }
\label{fig:folds}
\end{center}
\end{figure*}

Through the work of a number of teams, there are now some 105 believable planet candidates with periods between 4.2 hrs and 1 day (L\'eger et al.~2009; Winn et al.~2010; Batalha et al.~2011; Muirhead et al.~2012; Rappaport et al.~2012; Ofir \& Dreizler 2013; Huang et al.~2013; Sanchis-Ojeda et al.~2013a; Rappaport et al.~2013a; Jackson et al.~2013; and summarized by Sanchis-Ojeda et al.~2013b).  
Thus far, all planets found with $P_{\rm orb} \lesssim$ 3/4 day are small, i.e., $\lesssim 2\,R_\oplus$ and potentially rocky.  There are a number of reasons why larger planets cannot easily survive in such short-period orbits. Among the pitfalls of being a short-period gas giant are tidally-induced orbital decay (Rasio et al.\ 1996; Schlaufman et al.~2010), a possible tidal-inflation instability (Gu et al.~2003a), Roche-lobe overflow (Gu et al.~2003b; Ford \& Rasio 2006), and evaporation (Murray-Clay et al.~2009).  An Earth-mass rocky planet would be less susceptible to these effects, and in particular the rocky portion of the planet could survive evaporation nearly indefinitely (Perez-Becker \& Chiang 2013). In this regard, several hot-Jupiters have been observed to be losing mass via gaseous emissions, e.g., HD 209458b (Vidal-Madjar et al.~2003; Yelle 2004; Garcia-Munoz 2007; Murray-Clay et al.~2009; Linsky et al.~2010) and HD 189733b (Lecavelier des Etangs et al.~2010).  

Thus far, only one exoplanet has shown evidence for mass loss via dusty effluents -- KIC 1255b (Rappaport et al.~2012).  The size of the underlying planet is inferred to be quite small (perhaps nearly lunar size) because of the theoretical difficulty of removing dust grains from planets with a substantial gravity (Perez-Becker \& Chiang 2013).

In this work we report on a second planet candidate that appears to also be emitting dusty material -- KOI-2700b.  This object is in a 21.84 hour orbit about a mid-K host star (KIC 8639908).  The presence of a dust tail is inferred via a distinctly asymmetric transit profile.  In Section 2 we present the observational evidence for a dust tail in KOI-2700 using 16 quarters of long-cadence, and 2 quarters of short-cadence, {\em Kepler} data.  The asymmetry in the transit is apparent by eye, but we fit the transit to a mathematical (i.e., non-physical) model and quantify the degree of asymmetry.  In this same section we also describe a number of tests and sanity checks that we have performed on the data.  In Section \ref{sec:psf} we present a new method for extracting the flux time series from point-spread-function (``PSF'') fitting of the images at the pixel level.  These results are used to confirm our findings with the more conventional {\em Kepler} data products.  A semi-physical dust-tail model is developed and utilized in Section \ref{sec:model} to fit the transit. This section includes a description of the dust-tail trajectories, a discussion of the scattering and absorption cross sections for the likely dust material, and a calculation of the values of $\beta$, the ratio of radiation-pressure forces to gravity, that are likely to obtain in the vicinity of KOI-2700.  The fitted transit parameters are discussed at the end of that section, with emphasis on a few of the parameters that have significant physical interpretation.  Finally, we summarize our results and draw a number of conclusions about this system in Section \ref{sec:discuss}.

\section{Transits of KOI-2700}
\label{sec:light_curve}

\subsection{Data retrieval and preparation}
\label{sec:data}

\noindent
We utilized all 16 quarters of the {\em Kepler} long-cadence data as well as the two existing quarters (Q14 and Q15) of short-cadence data for KOI-2700.  Most of the light curves used for this study have been produced with the simple aperture photometry (SAP\!\_FLUX) data. 
The time series from each quarter was divided by its median value. Then, the normalized data from all quarters were stitched together into a single file.

Stellar variability and instrumental signals were filtered out as follows.  For each data point in the flux time series, $f(t_j)$, a linear function 
of time was fit to all the out-of-transit data points at times $t$ satisfying $|t -t_j| < P_{\rm orb}/2$, where $P_{\rm orb}$ is the 
orbital period. Then $f(t_j)$ was replaced by $f(t_j) - f_{\rm fit}(t_j) + 1$, where $f_{\rm fit}$ is the best-fitting function. This
approach is quite efficient at filtering out any variability on timescales longer than the orbital period, while hardly affecting the
transit profile.

To determine the orbital period we first utilized a Lomb-Scargle transform with the frequencies oversampled by a factor of 20.  We then fit a linear function to the peak of the base frequency and the peaks of the next twelve higher harmonics.  The best determined period is $P_{\rm orb} = 0.910023 \pm (4 \times 10^{-6})$ days.  From this same transform, but before filtering, we determined that the rotation period of KOI-2700 is $P_{\rm rot} = 10.98 \pm 0.02$ days.  We also used a box least squares transform (``BLS''; Kov\'acs et al.~2002) to search for other planets in the system, but could set only limits on the transit depths of $\sim$$100\,(P_{\rm orb}/{\rm days})^{1/3}$ ppm. 

We also determined the orbital period via a different approach.  Here, we carried out an oversampled grid search of periods via epoch folding and measuring $\chi^2$ for the hypothesis that there are no systematic deviations from a constant flux in the folded light curve.  Values of $\chi^2$ are mapped out vs.~trial period and the maximum $\chi^2$ determines the most likely period;  the shape of the $\chi^2$ profile and the number of bins in the fold determine the uncertainty in period.  The best-fit period via the trial folds is 0.910022 $\pm$ ($5 \times 10^{-6}$) days.

\begin{figure*}
\begin{center}
\includegraphics[width=0.47 \textwidth]{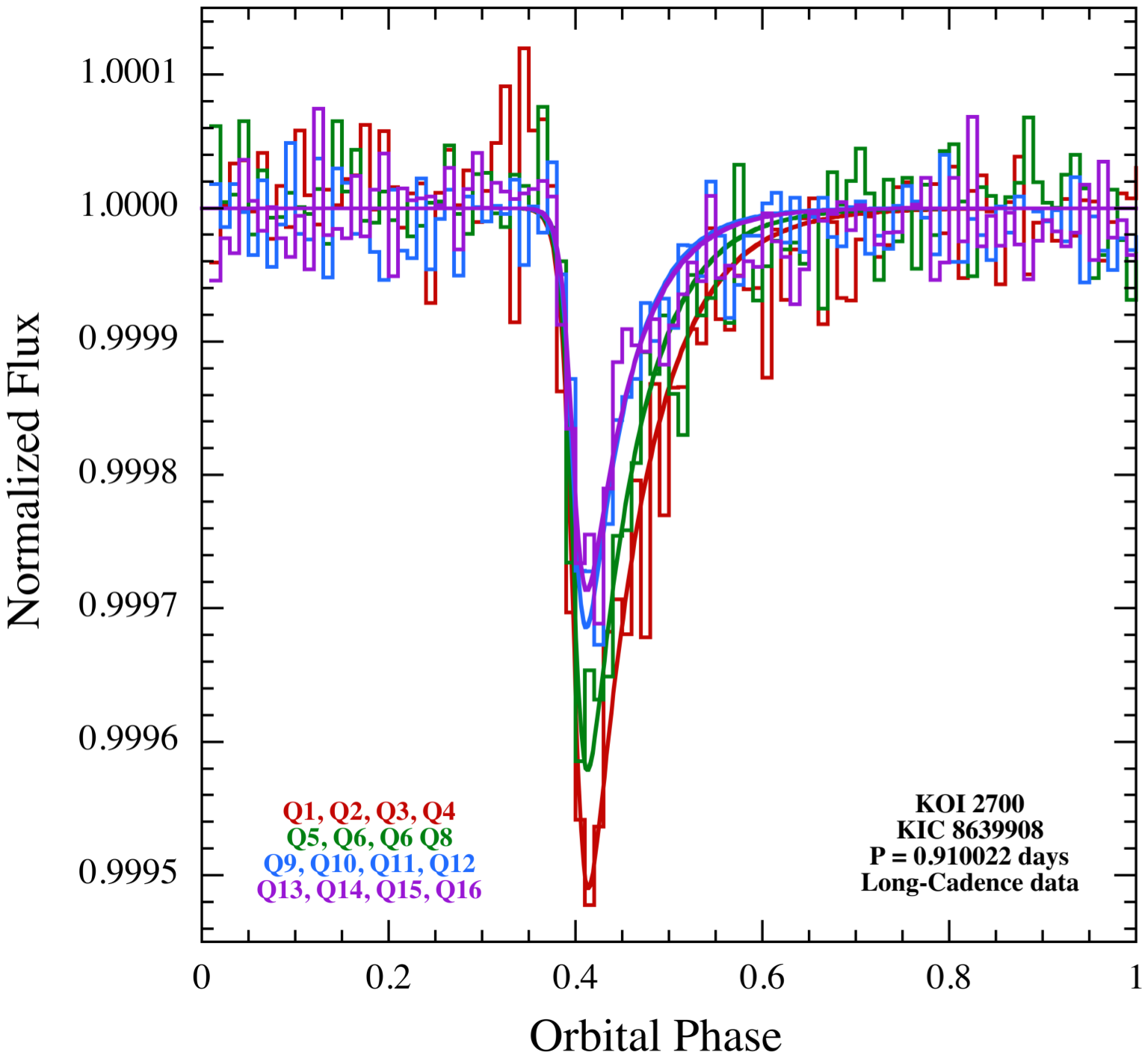}  \hglue0.3cm
\includegraphics[width=0.47 \textwidth]{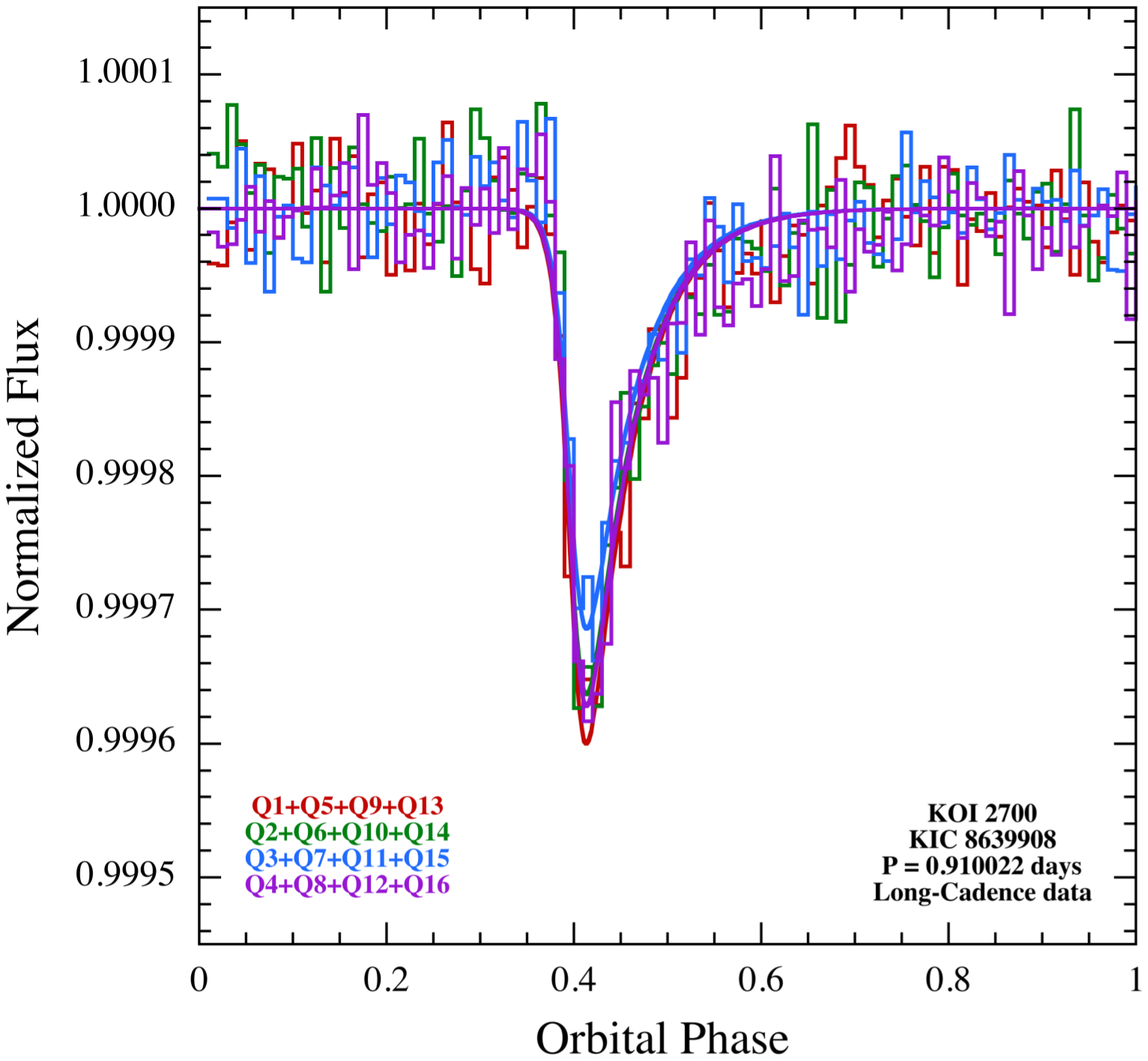}
\caption{{\it Left}.---Folded light curves for KOI-2700b broken up by years elapsed into the {\em Kepler} mission.  Note the dramatic decrease in transit depth with time.  
{\it Right.}---Folded light curves of KOI-2700b divided according to {\em Kepler} season (i.e., the spacecraft rotation angle).  Both the transit depth and shape are roughly independent of season.}
\label{fig:year_season}
\end{center}
\end{figure*}

\subsection{Average Light Curves}
\label{sec:LCs}

The 16 quarters of long-cadence filtered data for KOI-2700 were epoch-folded about the best determined period of $P_{\rm orb} = 0.910022$ days.  The folded data were then averaged into 200 bins per orbit, corresponding to a bin width of 6.5 minutes.  The results are shown in Fig.~\ref{fig:folds} (left panel).  The transit profile is obviously quite asymmetric, with a sharp ingress and much longer egress tail.  The corresponding light curve produced from the two available quarters of short-cadence data are shown in the right panel of Fig.~\ref{fig:folds}.  Because of the lower statistical precision of the short-cadence data (by a factor of $\sim$$\sqrt{8}$ due to the $\times$8 shorter baseline) we put the folded data into only 100 phase bins.  In spite of the reduced statistical precision of the short-cadence data, the same asymmetry of the transit profile is quite clear in the SC data.  

For purposes of quantifying the depth and asymmetry of the transit profile, we fit a non-physical mathematical function to the transit profiles.  The function we adopted is an ``asymmetric hyperbolic secant" (see, e.g., Ruan et al.~2000):
\begin{eqnarray}
T(\theta) = F_0 - C \left[e^{-(\theta-\theta_0)/\theta_1} +e^{(\theta-\theta_0)/\theta_2} \right]^{-1}
\label{eqn:AHS}
\end{eqnarray}
which has 5 free parameters: $F_0$ the out-of-transit flux, $C$ is related to the transit depth, $\theta_0$ is a measure of orbital phase zero, and $\theta_1$ and $\theta_2$ are the characteristic angular durations of the ingress and egress, respectively.  The minimum of this function occurs near $\theta_0$ but with an offset and a depth given by:  
$$ {\rm offset} = \theta_1\theta_2 \ln(\theta_2/\theta_1)/(\theta_1+\theta_2) $$ 
$${\rm depth} = \frac{  C \mathcal{R}^{  \mathcal{R}/(1+\mathcal{R} )  } } { (1+\mathcal{R}) } $$
where $\mathcal{R} \equiv \theta_2/\theta_1$.  In the limiting case where $\theta_2 = \theta_1$ this function reduces to a conventional hyperbolic secant with a offset equal to 0 and a depth of $C/2$.

Applying this fit to the long-cadence data yields $\theta_1 = (7.0 \pm 0.7) \times 10^{-3}$ and $\theta_2 = (6.6 \pm 0.35) \times 10^{-2}$, where the $\theta$'s are expressed in units of orbital phases.  The ratio of $\theta_2/\theta_1$, a direct measure of the asymmetry of the transit, is $9 \pm 1$; but, note that this value is really only a lower limit since the transit ingress is not well resolved with the LC data. The fitted value of the constant, $C$, plus the ratio $\mathcal{R}$, translate to a mean transit depth of $360 \pm 15$ ppm (over the four years of the {\em Kepler} mission).

\subsection{Variations with {\em Kepler} season and with time}
\label{sec:tests}

The relatively small transit depth for KOI-2700b does not readily permit an analysis of the transit-to-transit variability as could be done for KIC 1255b (see, e.g., Rappaport et al.~2012; Kawahara et al.~2013).  However, it is straightforward to check if there is a systematic variation of the transit profile or depth with time.  To this end, we divided the {\em Kepler} long-cadence data up into four approximately equal segments of 4 quarters each.  The results are shown in Fig.~\ref{fig:year_season} (left panel).  As is clearly evident, the transit depth is decreasing systematically with time over the duration of the {\em Kepler} mission.  The transit depths are: 525, 449, 305, and 308 $\pm$ 25 ppm, respectively, decreasing with year during the {\em Kepler} mission. Moreover, $\theta_2$, the ``recovery angle'' for the tail of the egress generally becomes systematically smaller with year (see Table \ref{tab:AHSfits}).

We also constructed epoch-folded light curves of KOI-2700b according to the four {\em Kepler} ``seasons'', i.e., by the rotation angle of the spacecraft.  These results are shown in Fig.~\ref{fig:year_season} (right panel).  We see that there are no obvious changes in either the shape or the depth of the transits with {\em Kepler} season (see Table \ref{tab:AHSfits}). We actually {\em do} expect to see a small change of $\sim$50 ppm from season 0 to season 4, separated on average by 9 months.  However, this change is approximately within the statistical uncertainty of the measurements. 

\begin{deluxetable}{lcccc}
\tablewidth{0pt}
\tabletypesize{\scriptsize}
\tablecaption{\label{tab:AHSfits} {Parameters Fits to AHS\tablenotemark{a} Model}}
\tablehead{
	\colhead{Epoch} & 
	\colhead{$C\tablenotemark{b}$} &
	\colhead{$\theta_1\tablenotemark{c}$} & 
      	\colhead{$\theta_2\tablenotemark{d}$} &
	\colhead{Depth\tablenotemark{e}}  \\
}
\startdata 
LC data & 493 & 0.0070 & 0.0655 & 360 \\
SC data & 419 & 0.0070 & 0.0860 & 321 \\
& & & & \\
year 1\tablenotemark{f}	& 716 & 0.0070 & 0.0669 & 525 \\
year 2\tablenotemark{f}	& 625 & 0.0070 & 0.0613 & 449  \\
year 3\tablenotemark{f} 	& 465 & 0.0070 & 0.0400 & 305  \\ 
year 4\tablenotemark{f}	& 452 & 0.0070 & 0.0473 & 308  \\
& & & & \\
season 0\tablenotemark{g} & 627 & 0.0070 & 0.0532 & 438  \\
season 1\tablenotemark{g} & 508 & 0.0070 & 0.0547 & 357 \\
season 2\tablenotemark{g} & 555 & 0.0070 & 0.0603 & 397  \\
season 3\tablenotemark{g} & 565 & 0.0070 & 0.0537 & 395 \\
& & & & \\
PSF fluxes & 471 & 0.0068 & 0.0598 & 339
\enddata
\tablenotetext{a}{Non-physical mathematical fitting function we have called an ``asymmetric hyperbolic secant'' (see also Ruan et al.~2000)}
\tablenotetext{b}{See Eqn.~\ref{eqn:AHS} for definition. Units are in ppm. Typical uncertainties are 8\%.}
\tablenotetext{c}{Fixed at the value found from the fit to all 16 quarters of the LC data. The egress is too steep to be resolved in the LC data, and there is insufficient statistical precision in the SC data to improve upon the determination of $\theta_1$.  For the PSF data set, $\theta_1$ was allowed to be a free parameter. Units are in orbital cycles.}
\tablenotetext{d}{Angle characterizing the egress tail; typical uncertainty of $\sim$6-10\%. Units are in orbital cycles.}
\tablenotetext{e}{Transit depth expressed in ppm; typical uncertainty of $\sim$8\%.}
\tablenotetext{f}{Year of the {\em Kepler} mission}
\tablenotetext{g}{{\em Kepler} spacecraft rotational season}
\end{deluxetable}

\begin{figure*}
\begin{center}
\includegraphics[width=0.48 \textwidth]{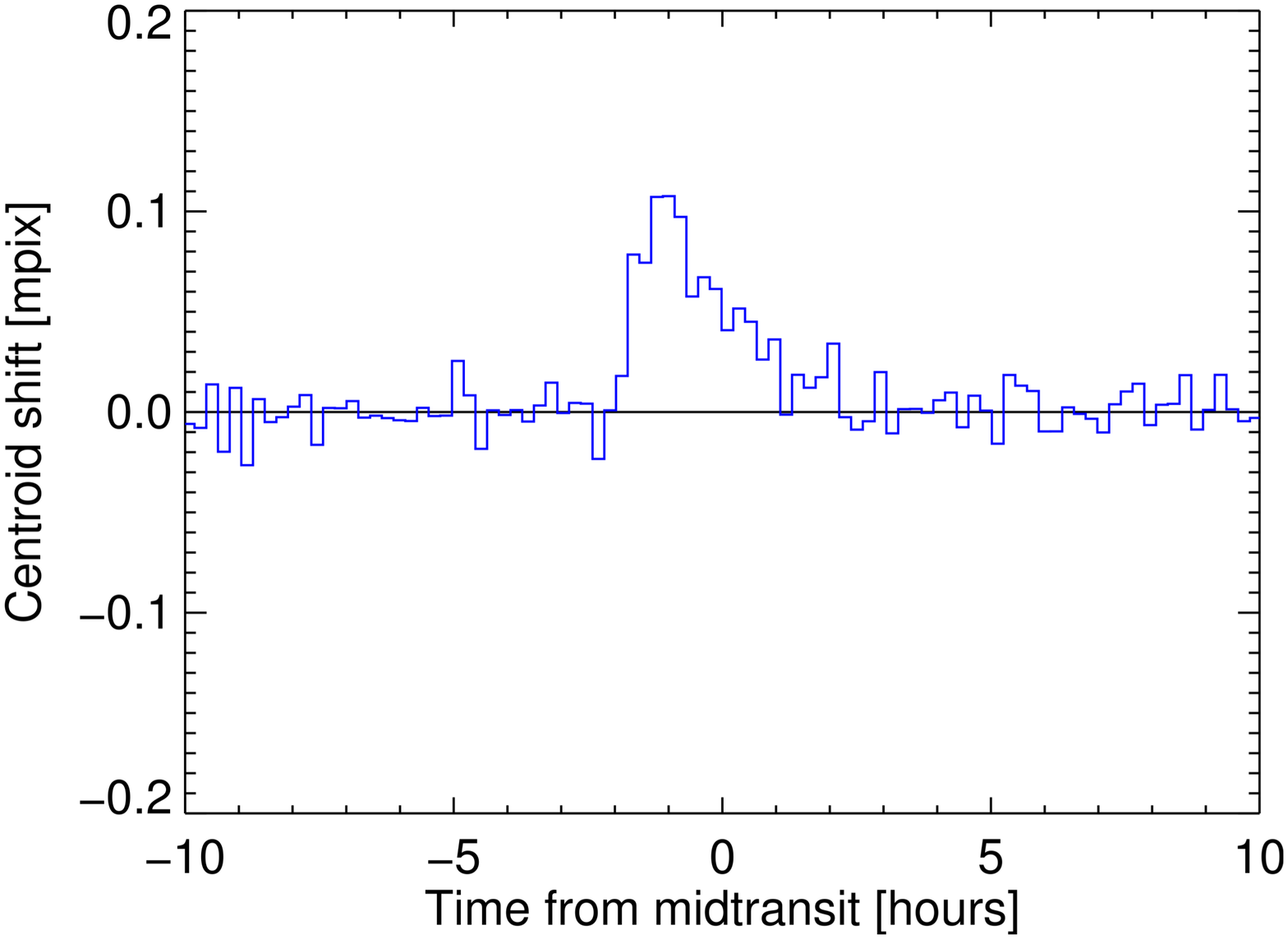}  \hglue0.1cm 
\includegraphics[width=0.48 \textwidth]{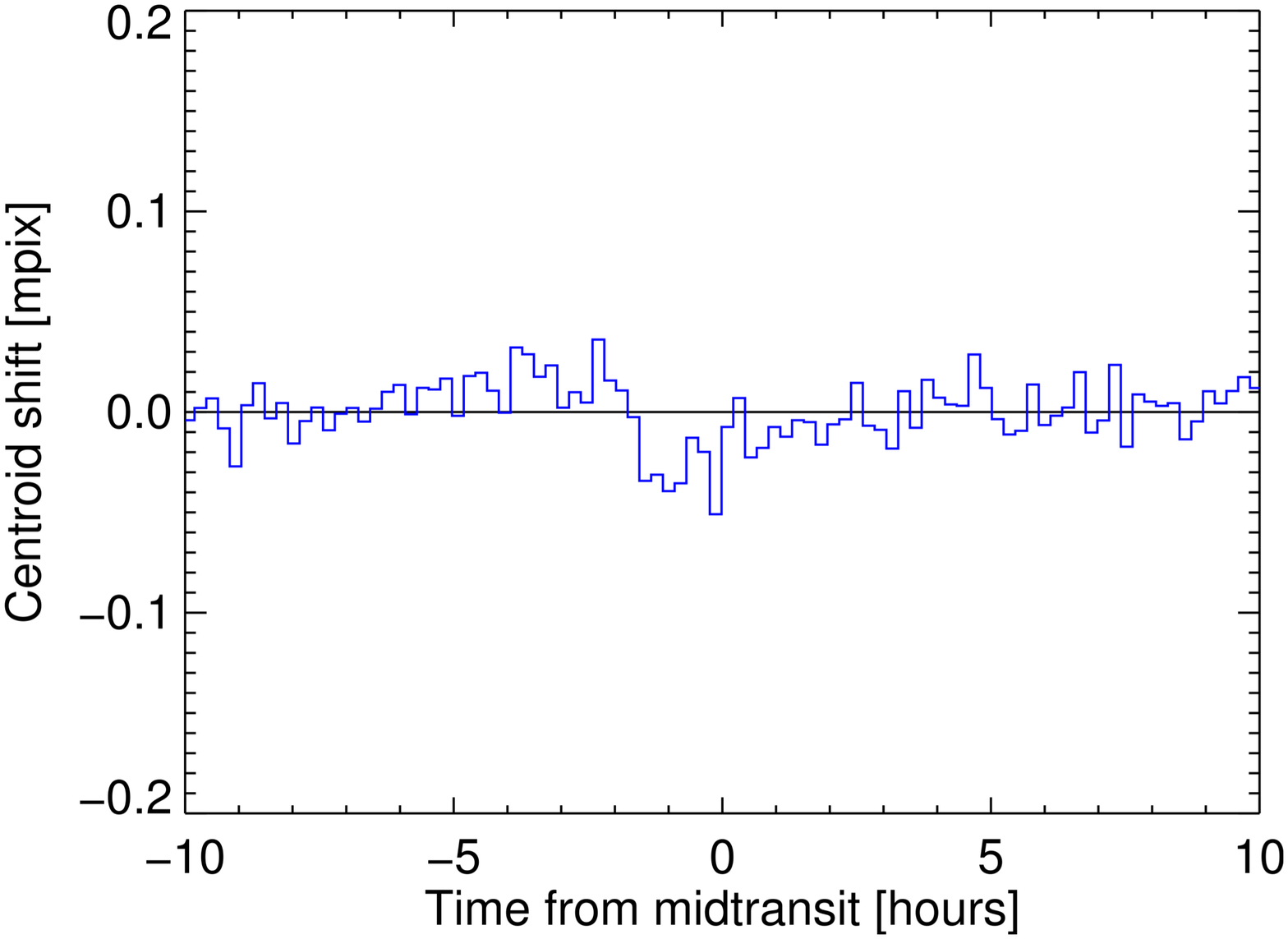}
\caption{{\it Left}.--- Image centroid shift of the column position for KOI-2700b folded with the period of 0.910022 days.    The $\sim$110  micropixel shift is due to the presence of a neighboring star that is 6.6$''$ away from the target star, but $\sim$ 2 magnitudes brighter. {\it Right}.--- Image centroid shift of the row position for KOI-2700b exhibiting a corresponding $-40$  micropixel shift. }
\label{fig:centroid}
\end{center}
\end{figure*}

The transit parameters found from the asymmetric hyperbolic secant fit are summarized in Table \ref{tab:AHSfits} for all four years and all four {\rm Kepler} seasons, as well as the summed 16 quarters of LC and 2 quarters of SC data.  

\subsection{Possible sources of light contamination}
\label{sec:3rd_light}

As is customary, we checked the image light centroid of the target star KIC 8639908 in phase with the transits in order to ascertain whether the source of the dips is indeed the target star (Jenkins et al.~2010).  We filtered the time series of the photocenter row and column pixel coordinates in the same manner that was described in Sect.~\ref{sec:data} for the flux time series.  We then calculated the mean of the in-transit coordinate, the mean of the out-of-transit coordinate, and the differences between those means, which we denote $dx$ and $dy$.  Fig.~\ref{fig:centroid} (left panel) shows the centroid data for the column position folded with the 0.910022-day period.  It is clear that the light centroid shifts in the positive column direction by an amount $dx \simeq 110$ micro-pixels during the times of the transits compared to the out-of-transit times (Fig.~\ref{fig:centroid}; left panel).  There is also a smaller shift of the light centroid in the row direction of $dy \simeq -40$ micro-pixels (Fig.~\ref{fig:centroid}; right panel).  The mean transit depth is $df \simeq 360$ ppm.  We then examined the ratios $dx/df$ and $dy/df$.  When these are multiplied by the pixel size (4$''$), one obtains the angular offset between the source of the transits and the out-of-transit image photocenter (see, e.g., Jenkins et al.~2010).  The magnitude of the angular offset is $\sim$$1.3''$. One can then compare this offset during transits to the locations of the stars revealed in the electronically available UKIRT images of the {\em Kepler} field.

\begin{figure*}
\begin{center}
\includegraphics[width=0.43 \textwidth]{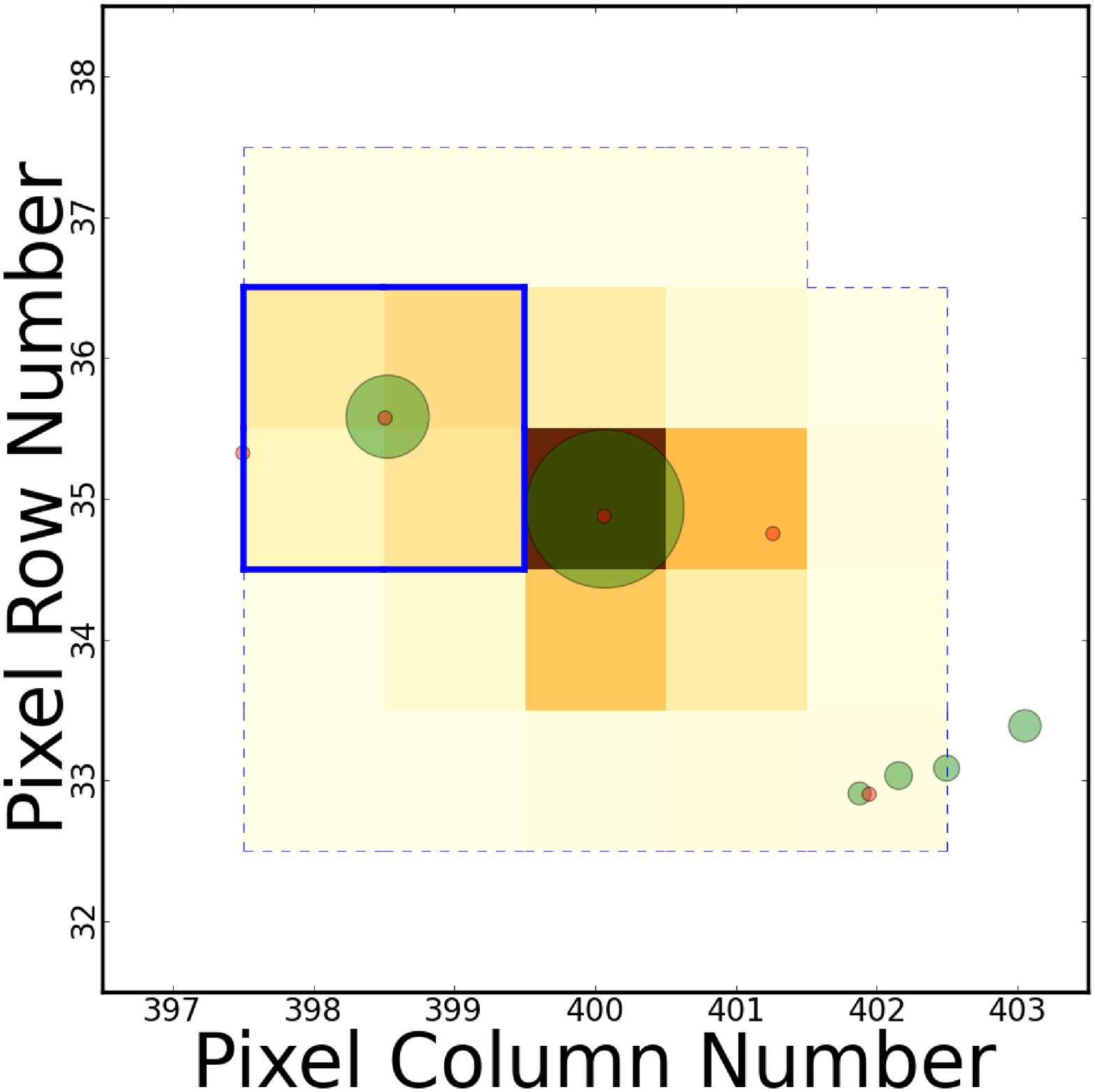}  \hglue0.1cm \includegraphics[width=0.52 \textwidth]{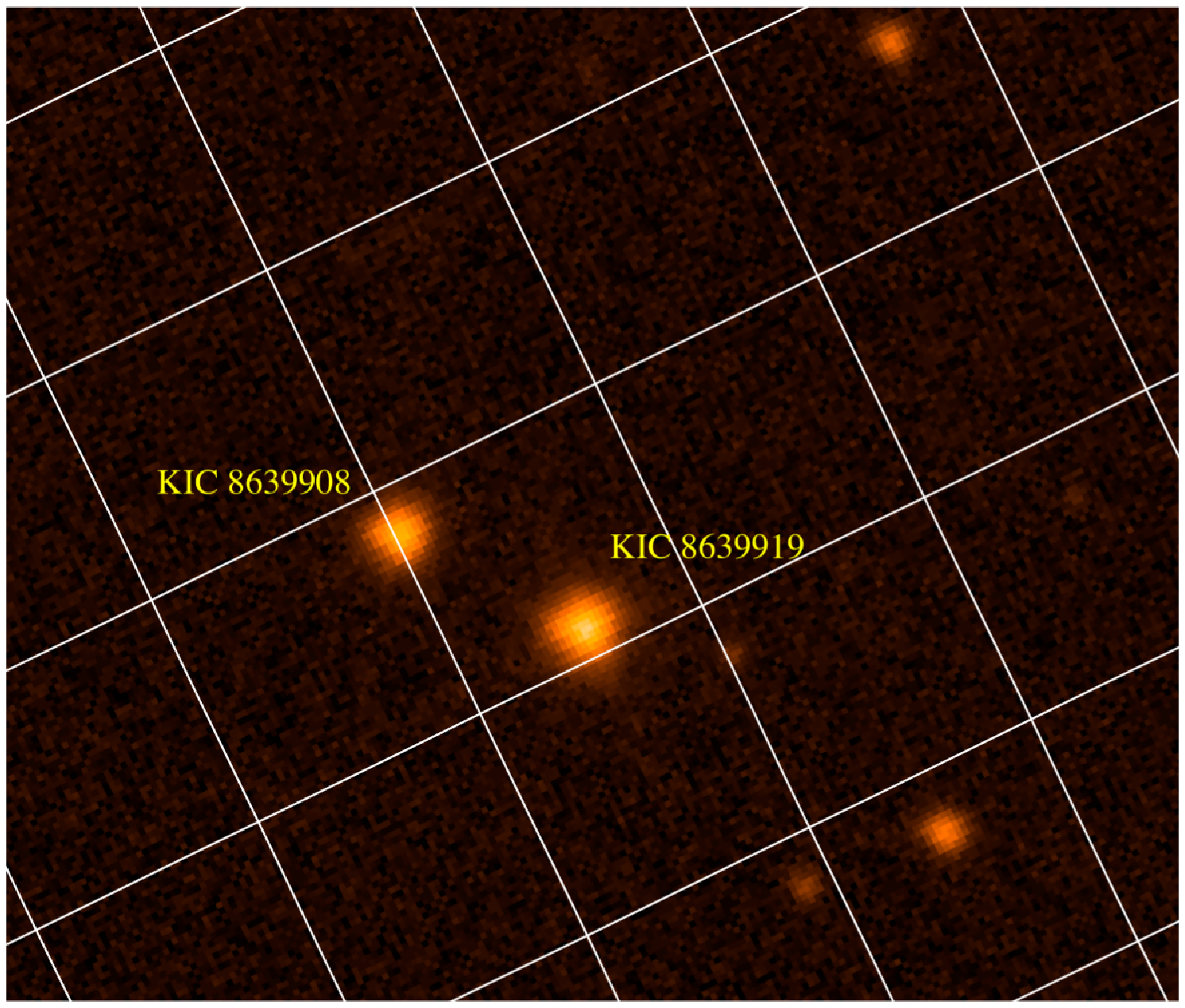}
\caption{{\it Left}.---An illustrative observational schematic at the pixel level from Q4. The dashed line contains the pixel mask for KIC 8639919, while the solid line is an $8'' \times 8''$ photometric aperture centered on KIC 8639908.  This pixel mask is actually for the brighter neighboring object at  \{400, 35\} that is $\sim$6.6$''$ to the southeast of KIC 8639908 at  \{398.5, 35.7\}; both objects happen to be contained in the same pixel mask.  Green circles are KIC sources, red circles are source detections from other catalogs.  The objective is to disentangle the fluxes from the target star and its neighbor which is $\sim$2 magnitudes brighter.  {\it Right.}---UKIRT J-band image of the KOI-2700 region, showing the target star and its brighter neighbor.  The orientation here is the same as for the schematic on the left.  
The grid lines are $8'' \times 8''$ and represent RA and Dec, with north pointing toward the lower left and east to the lower right.}
\label{fig:pixel_UKIRT}
\end{center}
\end{figure*}

\begin{center}
\begin{deluxetable}{lcc}
\tabletypesize{\scriptsize}
\tablecaption{Parameters of KOI-2700b and the Host Star\label{tbl:params}}
\tablewidth{0pt}
\label{tbl:parms}

\tablehead{
\colhead{Parameter} & \colhead{Value} & \colhead{Uncertainty} 
}

\startdata
Effective temperature, $T_{\textrm{eff}}$~[K]\tablenotemark{a}       &   4433  & $\pm 66$ \\ 
Mass of the star, $M_s$~[$M_{\odot}$]\tablenotemark{a}                  & $0.632$    & $\pm 10\%$  \\
Radius of the star, $R_s$~[$R_{\odot}$]\tablenotemark{a}        &  $0.574$  &  $\pm 10\%$ \\
Surface gravity, $\log g$~[cgs]\tablenotemark{a}  & 4.721 & 0.1 \\   
Metallicity, [Fe/H]\tablenotemark{a}  & $-0.2$ & ... \\
Stellar rotation period~[days]\tablenotemark{b} & 10.98 & $\pm 0.02$ \\ 
& & \\
Right Ascension (J2000)\tablenotemark{c} & 19$^h$ 00$^m$ 03$\fs$14 & ... \\
Declination (J2000)\tablenotemark{c} & 40$^\circ$ 13$'$ 14$\farcs$7& ... \\
$K_p$ mag\tablenotemark{c} & 15.38 & ... \\
$g$ mag\tablenotemark{c} & 16.44 & ... \\
$r$ mag\tablenotemark{c} & 15.30 & ... \\
$i$ mag\tablenotemark{c} & 14.93 & ... \\
$z$ mag\tablenotemark{c} & 14.69 & ... \\
$J$ mag\tablenotemark{c} & 13.58 & ... \\
$H$ mag\tablenotemark{c} & 12.99 & ... \\
$K_s$ mag\tablenotemark{c} & 12.84 & ... \\
& & \\
Reference epoch for folds~[BJD]        & 2454900.00  &  ...      \\
Orbital period~[days]                                 & $0.910022$  & $\pm 0.000005$    \\
Transit Depth [ppm]          & $305-525$\tablenotemark{d}         &  ...  \\
Time of Transit [BJD]  & 2454900.358 &  0.004 \\
Scaled semimajor axis, $a/R_s$                    & 5.9 & $\pm 0.4$  \\
Impact parameter, $b$                        & $<1.0$           &  ...    \\
Transit `duration' ~[hr]                 & $>5$   &   2 $\sigma$    \\
& & \\
Amplitude of ELV [ppm] & $<$\,10 &  2 $\sigma$ \\ 
& & \\
Planet radius, $R_p$~[$R_\oplus$]           &  $<$\,1.06         &  2 $\sigma$ \\
Planet mass, $M_p$~[$M_{\rm Jup}$]\tablenotemark{e}          & $<$\,0.86   &  2 $\sigma$ \\
& & \\
Sublimation constant, $S$\tablenotemark{f} [rad$^{-1}$] &   $2.4 \pm 1.0$	 & 2 $\sigma$ \\
$\beta$\tablenotemark{g}, $F_{\rm rad}/F_{\rm grav}$ & $<0.07$ & ... \\
\enddata
%\tablecomments{ }
\tablenotetext{a}{From Pinsonneault (2012). We assume a plausible uncertainty of 10\%.}
\tablenotetext{b}{Derived from the FT of the {\em Kepler} long-cadence data.}
\tablenotetext{c}{Taken from the Kepler Input Catalog, Batalha et al.~(2010).}
\tablenotetext{d}{Variable with time.} 
\tablenotetext{e}{Based on absence of ellipsoidal light variations, assuming zero dilution.}
\tablenotetext{f}{From the model fit -- see Sect.~\ref{sec:model}}
\tablenotetext{g}{Based on model calculations applicable to KOI-2700}
\end{deluxetable}
\end{center}

An inspection of the UKIRT J-band image of this region (see Fig~\ref{fig:pixel_UKIRT}; right panel) indicates that in addition to the J = 13.58 magnitude star at the location of the target star, KIC 8639908, there is a neighbor star (KIC 8639919), about 1 magnitude brighter in J-band, at a distance of 6.6$''$ to the southeast.  For the two stars in question, each has its {\em own} pixel mask and photometric aperture.  During each quarter, the photometric aperture for KIC 8639908 is chosen to optimize the signal of the target while minimizing the flux leakage from the brighter neighbor star (i.e., maximizing the S/N) in the production of the SAP\!\_FLUX time series.  

As an illustration of the pixel mask geometry, we show in Fig.~\ref{fig:pixel_UKIRT} (left panel) the pixel mask for the neighbor star KIC 8639919, which also contains KIC 8639908 (KOI-2700).  The geometry of that mask, in relation to the target and the neighbor star, is shown with the location of the various KIC sources indicated.      

The net centroid shift of $\sim$120 micro-pixels during the time of transits corresponds to a difference of $\sim$$1.3''$ between the light centroid and the source of the transits.  The direction of that vector points from the target star to the bright neighbor, thereby indicating that the target star, KOI-2700, is the likely source of the transits. 

\section{PSF-Photometry of the Data at the Pixel Level }
\label{sec:psf}

As a further check on the photometry of the target star and its brighter neighbor, and to verify which star is being transited, we also made use of the data at the pixel level.

There is some potential for the variable depth of the transits observed over time to be a systematic artifact of the archived SAP\!\_FLUX, and for systematic errors to occur in the absolute depth of the transits.  Artifacts would be a product of the source confusion within the photometric pixel aperture, especially if the wings of a bright contaminating neighbor star are comparable to, or even outshine, the target star. In SAP\!\_FLUX data, there is no attempt to correct for systematic flux deviations due to source crowding. To test the fidelity of the transit behavior inferred from the SAP\!\_FLUX data we employed a new technique to derive a complementary sequence of light curves by fitting a point-spread-function (i.e., ``PSF'') model to the calibrated pixel-level flux from each of 62,000 individual exposures across quarters Q1 to Q16. The empirically determined combined differential photometric precision (ÔCDPPÕ; Christiansen et al.~2012) of the PSF photometry for KOI-2700, with a median value of 147 ppm, is slightly inferior to the precision of the SAP\!\_FLUX data (median CDPP = 124 ppm) -- a consequence of the limitation in the precision of the PSF model. However we show that PSF photometry successfully deconvolves the two confused sources (KOI-2700 and its bright neighbor), verifies KOI-2700 as the transit source, confirms the observation of monotonically decreasing transit depths, and yields approximately the same average transit depth that is obtained from the SAP\!\_FLUX data from the pixel mask and photometric aperture for KIC 8639908. 

\subsection{The Basic PSF-FittingTechnique}

\begin{figure}
\begin{center}
\includegraphics[width=0.99 \columnwidth]{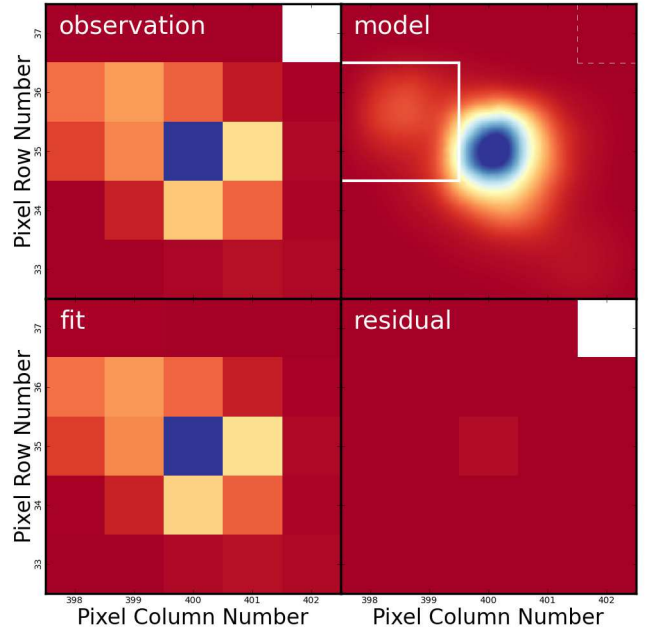}
\caption{Top-left is a typical 29.4-min.\,LC observation of the target, for one particular frame of 62,000 exposures. The blue pixel at \{400, 35\} contains KIC 8639919 that is $\sim$2 magnitudes brighter than the target KOI-2700 located at \{398.5, 35.7\}.  During this quarter, the target fell upon detector module 17 output node 3. The best PSF model fit is plotted in the top-right, while that fit, rebinned across the detector pixels, is compared at the lower-left. The lower-right panel contains the residuals between the data and the best fit.  All images are plotted on a linear intensity scale.  PSF-derived light curves are constructed by repeating this fit for every exposure over quarters 1 to 16.}
\label{fig:psf}
\end{center}
\end{figure}

With a number of caveats noted below, we assume that the pixels within the target mask of KOI-2700 can be characterized as two PSFs of independent magnitude and position (see Fig.~\ref{fig:psf}). We then minimize upon the uncertainty-weighted residuals between the data and the fit in order to yield the flux and pixel position of both sources for every exposure across Q1-16.  More specifically, there are two unknown shifts of the photometric aperture relative to the sky ($\Delta \alpha$ and $\Delta \delta$),  and two unknown source fluxes, that can be determined via a $\chi^2$ fit to the intensities in four or more pixels (see Fig.~\ref{fig:psf}).  

The {\em Kepler} PSF was calibrated from dithering observations during the commissioning phase of the mission (Bryson et al.~2010). The PSF is stored as a set of pixel-value lookup tables at the MAST archive\footnote{\url{http://archive.stsci.edu/kepler/fpc.html}}. Due to optical aberrations, the PSF varies in shape across the field-of-view. Each CCD channel has five PSF models associated with it - one for the center of the channel and one for each corner. While the models are of high statistical precision, their employment introduces two main sources of systematic error. First, the shape of the PSF must be interpolated to the target position from this relatively coarse grid of models. Second, telescope focus is temperature-dependent. The thermal state of the spacecraft changes as the sun-angle of the telescope boresight varies across the orbit and after spacecraft maneuvers. Therefore the accuracy of the commissioning calibration is a function of the thermal state of the spacecraft. 

For much more detail on how the PSF-fitting algorithm works and is implemented, see Still \& Barclay (2013).

\subsection{PSF-Photometry Results}

\begin{figure}
\begin{center}
\includegraphics[width=1.0 \columnwidth]{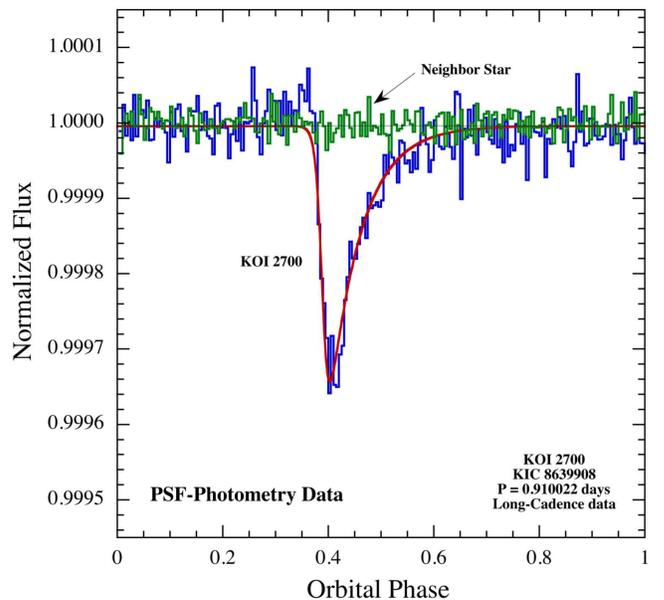}
\caption{Epoch-folded transit profile produced from PSF photometry at the pixel level for KIC-2700 (blue curve) and for the nearby bright neighbor star (green curve).  There is no trace of the transit in the photometry from the neighboring star.}
\label{fig:compare}
\end{center}
\end{figure}

We first utilized the PSF-photometry to demonstrate conclusively that the source of the transits is the target star, KOI-2700, and not the brighter neighbor star.  We had provided good, but indirect, evidence for this in Sect.~\ref{sec:3rd_light} above.  In Fig.~\ref{fig:compare} we show an epoch-folded light curve of the PSF-photometry for the target star KOI-2700.  The familiar asymmetric transit profile seen in the SAP\!\_FLUX photometry is nicely reproduced.  Superposed on this plot is the epoch-folded light curve using PSF-photometry for the bright neighbor star.  There is no trace of any transit.  This shows rather conclusively that we have identified the correct star that is being transited.

Second, we used the PSF-photometry to confirm, though with somewhat weaker statistical precision, that the transit depths are indeed {\em decreasing} with time during the {\em Kepler} mission.  These relative decreases in flux are in close agreement with those listed in Table \ref{tab:AHSfits} for the SAP\!\_FLUX data.

Finally, we note that the transit profile shown in Fig.~\ref{fig:compare} for KOI-2700, produced with the PSF-fitted flux series, has a transit depth of 340 ppm.  This is within the statistical uncertain of the mean transit depth obtained from the SAP\!\_FLUX data (see Table \ref{tab:AHSfits}).

Because the PSF-fitting technique is just being developed, we continue to utilize the results obtained with the SAP\!\_FLUX data for the remainder of this study, though the results from the two data sets are in excellent overall agreement.

\section{Model Fits}
\label{sec:model}

\begin{figure*}
\begin{center}
\includegraphics[width=0.47 \textwidth]{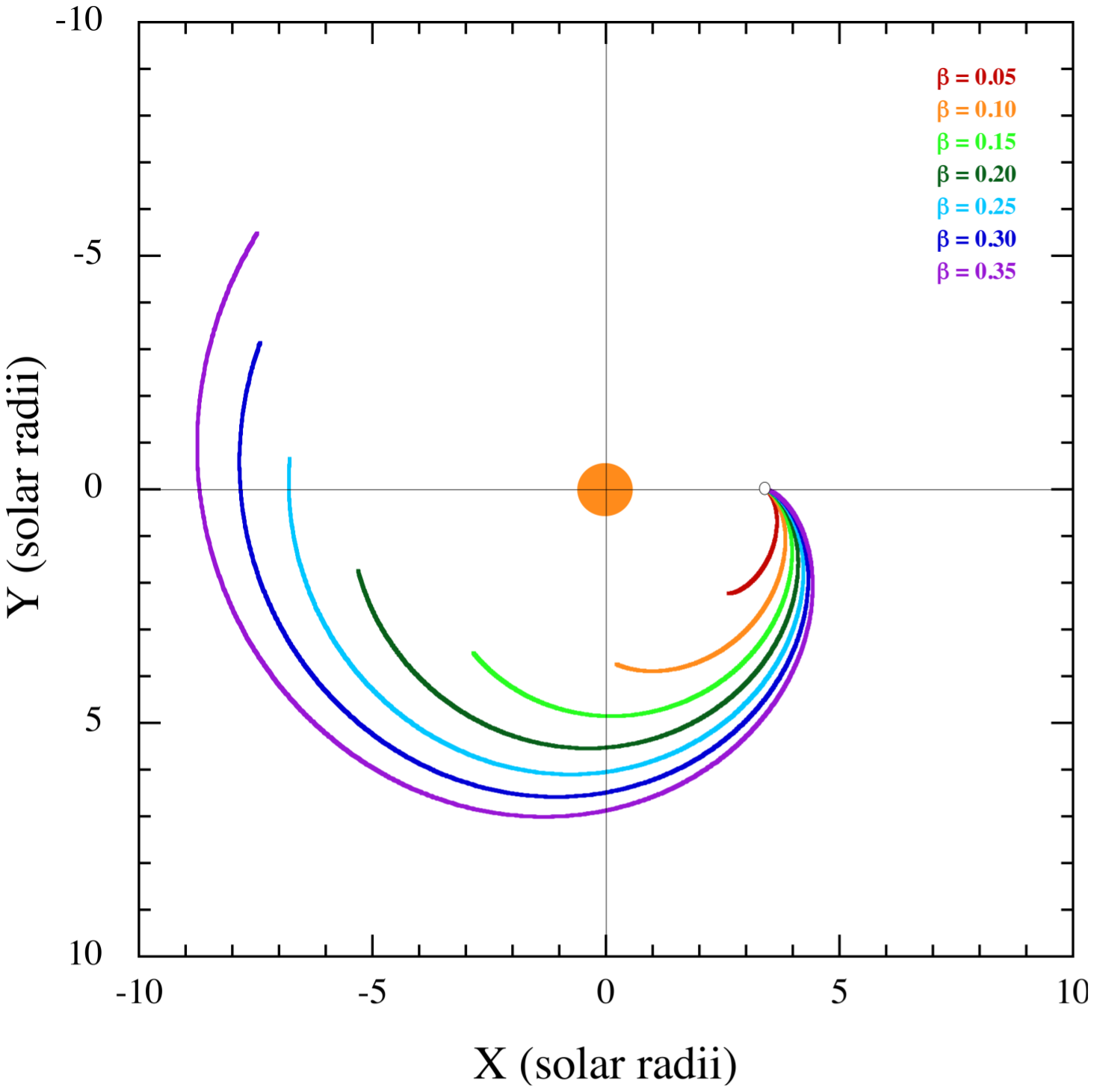} \hglue0.3cm
\includegraphics[width=0.478 \textwidth]{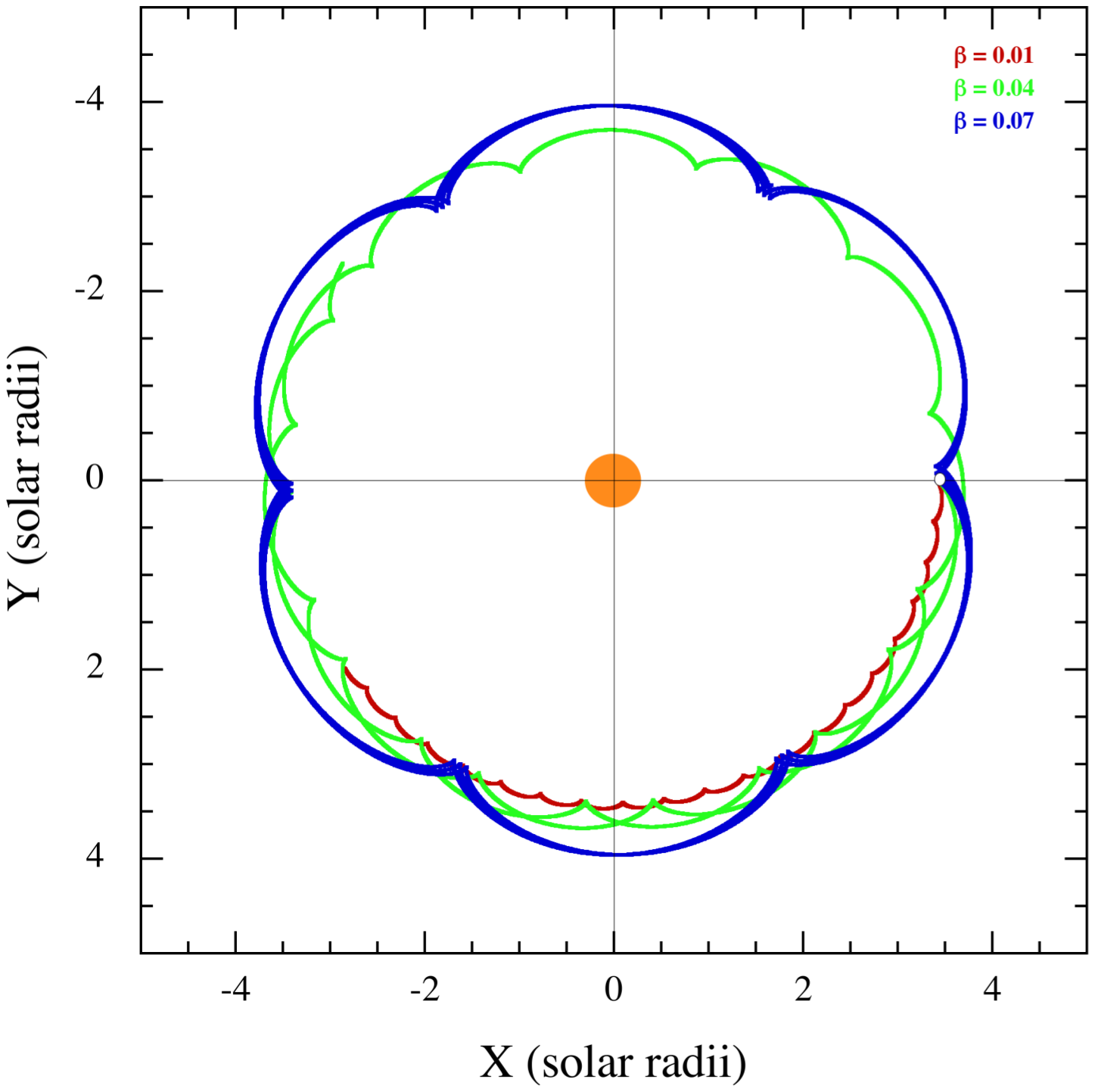}
\caption{{\it Left}.---Dust particle trajectories in the rest frame of the orbiting planet for different fixed values of $\beta$, the ratio of radiation pressure forces to gravity (see text).  The duration of the dust trail is arbitrarily cut off after one orbital cycle of KOI-2700b (i.e., a duration of 0.91 days).  The filled orange circle is supposed to represent the host star to approximately the correct scale..  
{\it Right.}---Same as the left panel, but for smaller values of $\beta$.  These dust particle trajectories are followed for 20 orbital periods of the planet. The cusps represent periastron passages of the dust particles where the instantaneous angular velocity of the dust equals that of the planet and the density would be enhanced.}
\label{fig:dust_tails}
\end{center}
\end{figure*}

We now proceed to analyze the asymmetric transit profile of KOI-2700b under the assumption that the transit is the result of a very small planet that is trailing a dust cloud which is largely responsible for the optical extinction.  This model is motivated by the dust tail that is inferred to be present in KIC 1255b (Rappaport et al.~2012; Brogi et al.~2012; Budaj 2013).  In any such comet-tail like model, the effective optical thickness of the dust tail is assumed to decrease monotonically with angular distance from the emitting planet.  The falloff in optical thickness is due to four possibilities: (i)  a decrease in the density of dust grains due to an apparent increase in angular velocity with distance from the planet (i.e., inferred from the continuity equation); (ii) a decrease in the diameter of the particles, and hence a concomitant decrease in cross section, as the particles sublimate;  (iii) unlikely geometric broadening of the dust tail perpendicular to the orbital plane so that it ultimately covers more than the diameter of the host star; and (iv) possible removal by stellar winds.  We discuss some effects of stellar winds separately in Appendix \ref{app:winds} and conclude that this is not likely to have a large effect on the dust tail (see, e.g., Cohen et al.~2011; Bisikalo et al.~2013 for a related discussion of the interaction of stellar winds with {\em gaseous} emissions from planets).

\subsection{Motion of the dust particles}
\label{sec:motion}

We can gain some considerable insight into the shape of the dust tail and the angular velocity vs.~angular distance by considering an idealized problem.  If the dust particles are ejected from the planet with a velocity that is comparable with the escape speed from the planet, then initially the liberated dust will be moving with basically the same orbital velocity as the planet.  The ratio of planet escape speed to orbital speed is $\sim$$\sqrt{M_p a_p/M_s R_p}$, where $M_s$ and $M_p$ are the masses of the host star and the planet, respectively, $R_p$ is the radius of the planet, and $a_p$ is the orbital radius of the planet.  If the mass ratio of star to planet is several hundred thousand, then this factor completely outweighs the small ratio of planet radius to orbital radius.  Thus, once the dust is free of the planet's gravity, it is essentially at rest with respect to the planet in its motion about the host star.

A liberated dust particle is then subjected to radiation pressure from the host star\footnote{For a brief discussion of the potential effects of a stellar wind on the dust see Appendix \ref{app:winds}.}.  For a given dust particle, the ratio of radiation-pressure force to gravity is designated as $\beta$.  Thus, ideally, the dust particle finds itself in an effectively reduced gravitational field compared to that of the planet, with $g_{\rm eff} = GM_s(1-\beta)/r^2$.  The resultant orbit of the dust particle is a Kepler ellipse with a periastron point at the location where it was released.  The eccentricity, $e$,  semimajor axis, $a_d$, and orbital angular frequency, $\omega_d$, of the dust particle are straightforward to work out.
\begin{eqnarray}
e = \frac{\beta}{1-\beta}; ~~a_d = a_p \frac{1-\beta}{1-2\beta};~~\omega_d = \omega_p \frac{(1-2\beta)^{3/2}}{1-\beta}
\end{eqnarray}
Moreover, the angular velocity of the dust particle with respect to the planet is given by a closed-form, analytic expression:
\begin{eqnarray}
\omega_p - \dot \theta_d = 2 \omega_p \beta \left(2-\beta+\beta \cos \theta_d \right) \sin^2(\theta_d/2)
\label{eqn:den}
\end{eqnarray}
where $\omega_p$ is the orbital frequency of the planet, and $\theta_d$ is the angular distance traveled by the dust particle in inertial space after its release.  If any quantities are needed as an explicit function of time, then it is straightforward to solve Kepler's equation.  

Some illustrative dust particle trajectories in the rest frame of the orbiting planet are shown in Fig.~\ref{fig:dust_tails} (left panel) for several different fixed values of $\beta$, the ratio of radiation-pressure force to gravity, over the range of $\beta = 0.05-0.35$ ($\beta > 0.5$ leads to an unbound orbit).  The duration of the dust trail is arbitrarily cut off after one orbital cycle of KOI-2700b (i.e., a duration of $\sim$22 hours).  Since the transit duration is at least 5-6 hours long, i.e., $\sim$25\% of the orbital cycle, this would require a value of $\beta \gtrsim 0.1$ (see Fig.~\ref{fig:dust_tails}) unless the dust can last for $\gtrsim$ 22 hours in the intense radiation field of the host star.

\begin{figure}
\begin{center}
\includegraphics[width=0.98 \columnwidth]{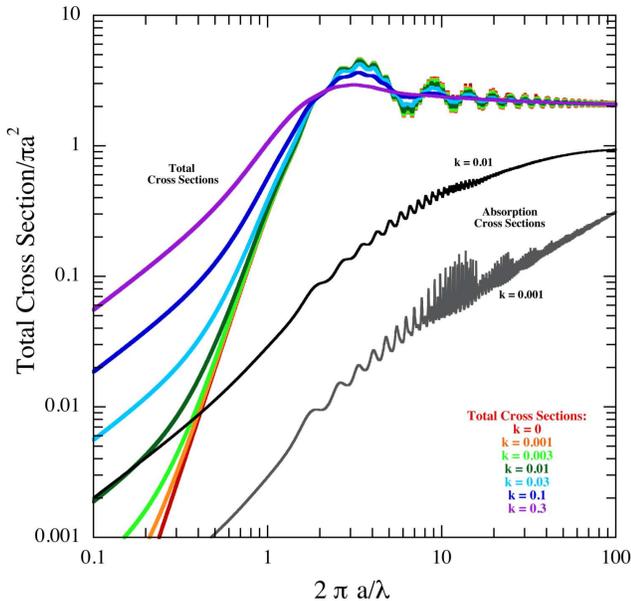} 
\caption{Mie scattering cross sections, normalized to the geometrical area of the particle, as a function of the particle size parameter, $2 \pi a/\lambda$, where $a$ is the particle radius and $\lambda$ is the wavelength of the radiation.  The real part of the index of refraction is taken to be $n_r = 1.65$, i.e., representative of materials from which the dust might be made, e.g., pyroxene.  For purposes of illustration, the imaginary part of the index of refraction, $k$, is taken to be a small constant, independent of $\lambda$.  Total cross sections are shown for 6 different values of $k$, while the absorption cross section is illustrated for two different values of $k$ ($k = 0.001$, grey curve; $k = 0.01$, black curve).  $k = 0.01$ is a reasonable upper limit to what might be expected for the imaginary part of the index at any relevant wavelength. }
\label{fig:sigma}
\end{center}
\end{figure}

We learned from simulations carried out for the study of KIC 1255b (Rappaport et al.~2012), that the dust particles remain in a rather tight ribbon (with respect to the orbital plane) as they stream away from the planet, nearly regardless of the value of $\beta$.  Thus, we expect from an argument involving the continuity equation, that the density of dust particles must necessarily thin out in proportion to $1/(\omega_p - \dot \theta_d)$ as the particles pick up angular speed with respect to the planet, assuming they are injected into the tail at a constant rate.  

Thus, we might hope that the inverse of Eqn.~\ref{eqn:den} would provide a good estimate of the density variation of dust particles along the tail (assuming there were no sublimation and no important effects from stellar winds).  If we wish to write this expression as a function of $\Delta \theta$, the angular distance between the dust particle and the planet, this requires the solution of Kepler's non-linear equation at every point.  However, an approximate transformation between $\theta_d$ and $\Delta \theta$ involves only the simple scaling relation:
\begin{eqnarray}
\theta_d ~& \simeq & ~\Delta \theta \frac{\omega_d}{\omega_{\rm syn}} = \Delta \theta \frac{P_p}{P_d-P_p} \\
& = & \Delta \theta \, \frac{(1-2\beta)^{3/2}}{1-\beta - (1-2\beta)^{3/2}} \simeq  \frac{\Delta \theta}{2\beta}
\end{eqnarray}
where $\omega_{\rm syn}$ is the synodic orbital period of the dust with respect to the planet: $\omega_{\rm syn} \equiv \omega_p - \omega_d$, and $P_d$ and $P_p$ are the orbital periods of the dust particle and planet, respectively.  We see that for small values of $\beta$ (see Sect.~\ref{sec:beta}) the factor multiplying $\Delta \theta$ can be quite large.  

It is apparent from the transit profiles that the angular extent of the dust tail exceeds 1-2 radians.  
Therefore, if $\Delta \theta$ in the dust tail where optical extinction is significant ranges from, say, 0 to 1.5 radians, then $\theta_d$ in Eqn.~\ref{eqn:den} could well range from 0 to 15 radians for an illustrative value of $\beta = 0.05$ (see Sect.~\ref{sec:beta}).  It would then follow that Eqn.~\ref{eqn:den} would cycle through several minima, with a concomitant set of maxima in the dust density.  However, when one takes into account a range of particle sizes in the dust emitted by the planet, this implies a corresponding range in values of beta (which depends sensitively on particle size; see Sect.~\ref{sec:beta}).  

As an example of what would ensue with a range of dust particle sizes, we show in Fig.~\ref{fig:dust_tails} (right panel) a set of dust particle trajectories in the rest frame of the planet, for three illustrative particle sizes: $a =\, $0.2, 0.1, and 0.05 $\mu$m, with corresponding values of $\beta \simeq \,$0.07, 0.04, and 0.01 (see Sect.~\ref{sec:beta}).

From the above discussion about the motion of particles of different sizes, we conclude that the periodic enhancements in density along the tail for a given size particle, will tend to be completely washed out by a distribution of particle sizes with their concomitant range of values of $\beta$.  Therefore, we will simply assume that the density of dust particles is approximately {\em constant} over the portion of the dust tail that is relevant to the observed asymmetric transit profile.  However, this does {\em not} mean that the attenuation of the dust is constant along the tail.  The particles will be subject to sublimation, and their effective cross section drops as their radii become smaller (see Sect.~\ref{sec:subl}); we return to discuss this in Sect.~\ref{sec:observ} and Appendix \ref{app:appA}.

\subsection{Estimation of $\beta$}
\label{sec:beta}

We approximate the dust particles as spherical and to have a generic real part of their indices of refraction, $n_r$, equal to 1.65.  We take the dust composition to be Earth-abundant refractory minerals such as pyroxenes and olivines.  The imaginary parts of the indices of refraction, $k$, for these materials typically tend to be small (compared to $n_r$) with relatively modest enhancements at specific wavelengths, $\lambda$, but with $k \lesssim 0.01$ at all $\lambda$ in the range of interest (Kimura et al.~2002, and references therein; Budaj 2013).  We utilize a Mie scattering code (Bohren \& Huffman 1983) to compute the various cross sections, $\sigma_s$ and $\sigma_a$ (scattering and absorption) and $\sigma_{\rm ext} \equiv \sigma_{\rm tot}$, the extinction or ``total'' cross section, for the dust particles as a function of their scaled size (i.e., $2 \pi a/\lambda$), where $a$ is the radius of the (assumed) spherical dust particles. The results for $\sigma_{\rm tot}$ are shown in Fig.~\ref{fig:sigma} for six different values of the imaginary part of the index of refraction, $k$ (assumed independent of $\lambda$, so as to render $\sigma/\pi a^2$ a function of the particle scale size only).  As the figure illustrates, the cross section is essentially twice the geometrical area for scale sizes $\gtrsim 2$, but falls roughly as a steep power law for small particle scale sizes.  The logarithmic slope, just below particle scale sizes of $\sim$ unity, depends sensitively on the value of $k$.  Two illustrative curves of $\sigma_a$ vs.~scaled particle size (for $k = 0.001$ and 0.01) are also shown in Fig.~\ref{fig:sigma} for comparison.  

\begin{figure}
\begin{center}
\includegraphics[width=0.98 \columnwidth]{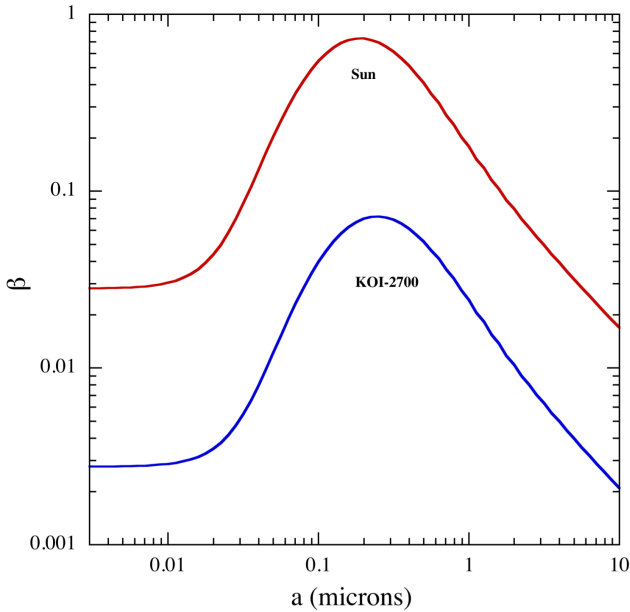} 
\caption{Calculated values of $\beta$, the ratio of radiation pressure forces to gravity, as a function of particle size, for an assumed composition of pyroxene (blue curve).  The indices of refraction used in the calculation are $n_r = 1.65$ and $k = 0.01$, i.e., the same values as were used to produce the green curve in Fig.~\ref{fig:sigma}. The luminosity and mass adopted to compute $\beta(a)$ in the vicinity of KOI-2700 are $L=0.09\,L_\odot$ and $M = 0.63 \, M_\odot$.  The red curve is the same function $\beta(a)$ in the vicinity of the Sun, shown for comparison.}
\label{fig:beta}
\end{center}
\end{figure}

We used the cross sections shown in Fig.~\ref{fig:sigma} to compute the value of $\beta$ as a function of particle size.  This requires averaging the cross sections over the radiation spectrum of the host star.  Given the level of approximations that has already been made (especially in view of the uncertainties in the chemical composition and shapes of the dust particles), we elected to utilize a simple blackbody spectrum with $T_{\rm eff} = 4433$ K (see Table \ref{tbl:params}).  The results for $\beta(a)$ are shown in Fig.~\ref{fig:beta}, with a comparison to $\beta(a)$ in the solar vicinity.  The peak value of $\beta$ occurs for particles of size $\sim$$0.24 \, \mu $m, with a maximum value of $\beta = 0.07\pm 0.02$ for KOI-2700 (where the error bar arises from the uncertainty in the stellar luminosity and mass).  This is in agreement with the results of Kimura et al.~(2002), after simply scaling their results for dust particles in the vicinity of the Sun (see their Fig.~2).  The scaling factor is just $(L_s/L_\odot)(M_\odot/M_s) \simeq 0.14 \pm 0.04$.  The Kimura et al.~(2002) maximum values of $\beta$ for pyroxene in the solar neighborhood are in the range 0.7-0.8.  Using the scaling factor above would yield $\beta_{\rm max} \simeq 0.06-0.14$ near KOI-2700, in fair agreement with our estimates.

\subsection{Sublimation of the dust}
\label{sec:subl}

A substantial difficulty with computing the transmission of the dust tail is that the dust particles will decrease in size monotonically with time due to sublimation in the intense radiation field.  As discussed in detail by Kimura et al.~(2002), the rate of sublimation depends on the instantaneous grain temperature, which in turn results from a thermal balance between heating and cooling.  As the dust particle grows smaller with its sublimation, the absorption cross section decreases, and the value of $\beta$ changes, as does its orbit.   

To obtain a simpler estimate of the sublimation lifetime for the dust particles in KOI-2700, we utilize the results of Kimura et al.~(2002) for the  lifetimes of cometary dust particles of different sizes and compositions as a function of heliocentric distance.  We scale the results in their Fig.~4 to the case of the KOI-2700 system.  The equivalent heliocentric distance at which to evaluate the grain lifetime is given by: $d_{\rm comet}/R_\odot = (L_\odot/L_s)^{1/2}(a_p/R_\odot) \simeq 11 \pm 2$.  At such distances from the Sun, amorphous olivine lasts for only a matter of minutes before it sublimates away; its crystalline counterpart could last hours or more at the nominal equivalent heliocentric distance of 11 $R_\odot$.  However, pyroxene grains (amorphous or crystalline) might well last for days even at the lower limit on equivalent distance of 9 $R_\odot$ before sublimating.

See Appendix \ref{app:appA} for a further discussion of grain sublimation.

\subsection{Transit Model}
\label{sec:observ}

For the transit model we require a function that gives (i) the number density of the dust particles as they thin out due to their changing velocity with respect to the planet, and (ii) the decreasing scattering cross section due to their sublimation.  For the first part, we concluded in Sect.~\ref{sec:motion} that, for small values of $\beta$ (appropriate to KOI-2700), use of an approximately constant number density of dust grains over a broad angular range in the dust tail is quite reasonable.  Calculation of the reduction in cross section due to sublimation is rather involved -- beyond the scope of this paper and, in fact, at present there may be insufficient information to compute it in a completely satisfactory way.  

Complications in the sublimation-induced changes in the dust-scattering cross sections include the fact that there is an unknown dust particle size distribution (``PSD'') emitted from the planet - and of unknown composition.  Furthermore, the particles begin to sublimate with elapsed time after their escape, and their radii change in a complex way.  The change in radius depends on the heating and cooling equation (see Appendix \ref{app:appA}), the imaginary part of the index of refraction, the orbit which, in turn, depends on $\beta$, and so forth.  The rates of change in particle size cause the PSD to vary with time, which in turn results in the optical thickness of the dust tail decreasing in a highly uncertain way.

In spite of these complications, we have carried out some simplified calculations of the rate at which particles decrease in size due to sublimation -- see Appendix \ref{app:appA}.  In the process, we have shown why an exponential decrease with time in the effective scattering cross section of the dust grains might be approximately correct. 

Thus, to represent the diminution of the grain sizes, and hence scattering cross section, with time and angular distance from the planet, we adopt a somewhat {\em ad hoc} exponential function with the e-folding angle as a free parameter (see also Rappaport et al.~2012; Brogi et al.~2012; Budaj 2013; Appendix \ref{app:appA} of this paper).  The specific functional form that we adopt to represent the attenuation of the dust tail for any angular distance, $\Delta \theta$ from the planet is:
\begin{eqnarray}
\mathcal{A}(\Delta \theta) = Ce^{-S\Delta \theta}
\label{eqn:tau}
\end{eqnarray}
where $\Delta \theta \equiv \theta - \theta_p$, and $\theta_p$ is the angular location of the planet, with $\theta_p = 0$ defined as the point where the planet is midway across the host star.  This function has only two free parameters: $C$, the normalization, and $S$, the exponential decay factor.

\begin{figure}
\begin{center}
\includegraphics[width=0.98 \columnwidth]{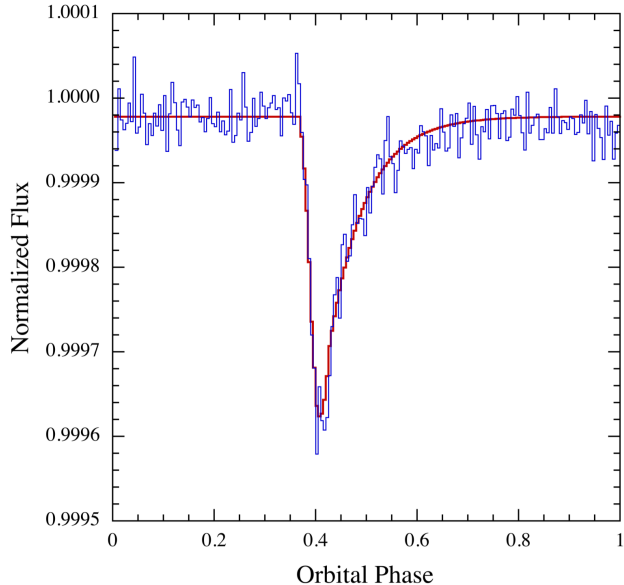} 
\caption{Illustrative model dust-tail fit to the folded long-cadence light curve for KOI-2700b for all 16 quarters (see text for model parameters). Each of the 200 bins is 6.5 minutes in duration, but the {\em Kepler} integration time is 29.4 minutes.  The model takes the finite LC integration time into account.  This particular model fit includes a hard-body planet with radius $R_p = 0.74\, R_\oplus$.}
\label{fig:model_fit}
\end{center}
\end{figure}

In order to use this transmission function to compute the transit profile, it must be {\em integrated} across a chord of the host star (for a given impact parameter, $b$) for each angular location of the planet around its orbit.  

The impact parameter, $b$, is a third free parameter, and the out-of-transit flux, $F_0$ is a fourth free parameter of the model. We also need to fit for the time (i.e., orbital phase) when $\theta_p = 0$.  This is the fifth, and possibly final, parameter in the transit model.  As an option, we can also include a hard-body planet attenuator, i.e., the contribution of a conventional transit for a planet of radius $R_p$.

Finally, we note that because there is no compelling evidence in KOI-2700b for enhanced forward scattering in the transit light curve (as is seen in KIC 1255b; Rappaport et al.~2012; Brogi et al.~2013; Budaj 2013), and due to the generally weaker transit signal in KOI-2700b, we do not include in the model preferential forward scattering into the observer's line of sight.  

\subsection{Fitted Parameters}

We used the above model to fit the transit profile for KOI-2700b.  The fits are generally quite good, with or without a hard-body planet included, and an illustrative result is shown in Fig.~\ref{fig:model_fit}.  Of the six free parameters in the fit, only three of them are of particular physical significance: $S$, the (assumed) exponential falloff with angle of the effective dust-tail cross section;
$b$, the impact parameter of the dust tail across the host star; and $R_p$, the hard-body planet radius.  The best-fit values are: $S = 2.4 \pm 1.0$ (in units of rad$^{-1}$), $b < 1.0$, and $R_p < 1.06 \, R_\oplus$ (90\% confidence limits).  

The range of values for $b$ is not constrained at all, as long as the dust tail crosses any part of the host.  The angular size of the star, as seen from the vantage point of the planet is small compared to the angular extent of the dust tail, and therefore the impact parameter has relatively little effect on the fit.  
The fitted value of the planet radius yields only an upper limit of $1.06 \, R_\oplus$.  The reason for the upper limit is the fact that a pure dust tail model fits the transit profile quite well, without the inclusion of a hard-body planet.

We can make use of the fitted transit model parameters to infer something about the lifetime of the dust particles after ejection from the planet.  The most probable value of the parameter $S$ is 2.4 rad$^{-1}$.  If we approximate the angle $\Delta \theta$ in Eqn.~\ref{eqn:tau} by $\Delta \theta \simeq \omega_{\rm syn} t$, where $t$ is the elapsed time for the dust particle to travel from the planet to its current location, then we can define a characteristic decay time for the cross section of the particles by
\begin{eqnarray}
\tau_d \simeq \frac{1}{S \omega_{\rm syn}} \simeq \frac{P_p}{2 \pi S} \,\frac{(1-\beta)}{(1-\beta) -(1-2\beta)^{3/2}} \simeq \frac{P_p}{4\pi S \beta}
\end{eqnarray}
This implies that the dust in KOI-2700 has a characteristic lifetime against sublimation of $\sim$$0.03/\beta$ days, where $\beta$ can have a range of values between $\sim$0.07 and $\sim$0.01.  Thus $\tau_d$ likely lies in the range of $\sim$0.5 to 3 days.  

\begin{figure}
\begin{center}
\includegraphics[width=0.99 \columnwidth]{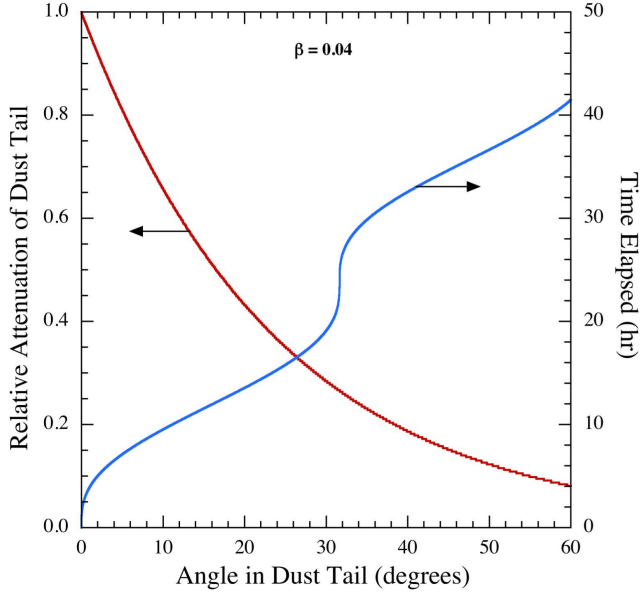} 
\caption{The dust-attenuation portion of the transit model (for an illustrative good fit) is plotted as a function of the angle, $\Delta \theta$, from the planet to the point in the dust tail contributing to the attenuation.  The corresponding interval of time required for a dust particle to reach an angular distance $\Delta \theta$, is also plotted.  The time was computed for the illustrative example of $\beta = 0.04$, i.e., the ratio of the radiation pressure to gravity.}
\label{fig:trans_vs_theta}
\end{center}
\end{figure}

We show in Fig.~\ref{fig:trans_vs_theta} an illustrative example of the dust attenuation contribution to the transit function (see Eqn.~\ref{eqn:tau}) vs.~the angular distance from the planet to a given point in the dust tail, $\Delta \theta$ (for the case where $S=2.4$).  The attenuation has fallen by a factor of $\sim$10 after one radian.  On the same plot, we show the time that is required for a dust particle to reach various angles, for an illustrative value of $\beta = 0.04$.  As can be seen from the figure, it takes $\sim$40 hours for dust to reach an angular distance of 1 radian from the planet.  Thus, at least some of the dust in the tail must last for approximately this long before sublimating.  

\section{Summary and Conclusions}
\label{sec:discuss}

In this work we have reported on the discovery of a second planet candidate, KOI-2700b, that likely exhibits a dust tail.  The optical attenuation of the dust tail appears to decrease monotonically with time during the {\em Kepler} mission with transit depths ranging between 525 and 305 ppm.  This likely implies a decrease in the rate of injection of dusty effluents into the tail by a factor of $\sim$2 over four years.  The magnitude of KOI-2700 ($K_p = 15.4$) and the relatively shallow average transit depth ($\sim$360 ppm)  makes it difficult to study possible orbit-to-orbit fluctuations in the transit depth, as was done in the case of KIC 1255b.  

The properties of the host star, KOI-2700 (see Table \ref{tbl:params}) are quite close to those of the other dust-emitting planet, KIC 1255b.  However, there are other planet candidates with $P_{\rm orb} \lesssim 1$ day and with similar host stars (Sanchis-Ojeda 2013b) which do {\em not} exhibit asymmetric transit profiles.  The main difference between KOI-2700b and KIC 1255b, in comparison with other short-period rocky planets, may be the {\em planet mass}.  As discussed in Rappaport et al.~(2012) and Perez-Becker \& Chiang (2013), it is extremely difficult (on theoretical grounds) to remove dust from planets with substantial gravity, e.g., Earth-sized.  In particular, Perez-Becker \& Chiang (2013) concluded that KIC 1255b is lower in mass than Mercury, and may be as small as the Moon.  Our upper limit on the hard body planet in KOI-2700 is $\sim$$1.06 \, R_\oplus$, consistent with, but not yet limiting this object to lunar size.  

The lifetime of such a small planet can be quite short if it loses an Earth mass per Gyr (see, e.g., Rappaport et al.~2012; Perez-Becker \& Chiang 2013).  However, since KOI-2700b has a transit depth that is, on average, an order of magnitude smaller than that of KIC 1255b, it may be losing mass an order of magnitude more slowly and therefore have a more palatable lifetime (i.e., $\gtrsim 100$ Myr).  

Because the transit is detectable for nearly a quarter of the 0.91-day orbital cycle, it implies an angular extent of some 90$^\circ$ around the host star.  For plausible values of the parameter $\beta$, the ratio of radiation-pressure forces and gravity, more than about a day is required for a typical dust particle to move 90$^\circ$ away from the planet.  Known abundant minerals that possibly can last this long in the intense radiation field of the host star (at a distance of $\sim$6 stellar radii) are pyroxine and crystalline olivine (see, e.g., Kimura et al.~2002).  Though, we note that Perez-Becker \& Chiang (2013) report that pyroxene is very difficult to remove from even sub-Mercury mass planets in the requisite quantities because of its much lower vapor pressure than olivine.  

At the current time, it appears that there are two major conditions for the generation of dusty comet-like tails on short-period planets.  The first was discussed by Rappaport et al.~(2012) and suggests that  the planet must lie in a favorable equilibrium temperature zone where, $T_{\rm eq} \equiv T_{\rm star}\sqrt{R_{\rm star}/a_{\rm orb}}$ is $\gtrsim 2000$ K.  Specifically, $T_{\rm eq}$ must be high enough to provide for a molten planet surface (a `lava ocean') with substantial vapor pressures of high-Z materials (i.e., O, Mg, Si, Al, Ca, Fe; see, e.g., Schaefer \& Fegley 2009, L\'eger et al.~2011), and the generation of an efficient Parker wind (Parker 1958) to drive material from the planet which then condenses into grains as the vapor cools (Perez-Becker \& Chiang 2013).  In the type of Parker wind envisioned here (see Perez-Becker \& Chiang 2013), the base of the atmosphere, which is heated by broadband radiation from the host star, drives a hydrodynamic bulk flow of matter which transitions from subsonic to supersonic as it escapes the planet.  If conditions are right, this mechanism can drive substantially more mass loss than the Jeans mechanism for atmospheric evaporation (Jeans 1904).  Once the dust has formed and is carried away from the planet, $T_{\rm eq}$ cannot be so high that these newly formed and liberated grains rapidly sublimate.  

The second condition, which is at least as important as the first, is the requirement that the planet have a sufficiently low surface gravity (or escape speed) so that it is possible for a Parker wind to develop at the atmospheric temperatures typical for close-in planets around K stars (see Rappaport et al.~2012; Perez-Becker \& Chiang 2013).  

As a crude estimate (explored somewhat further in Rappaport et al.~2012, and in detail in Perez-Becker \& Chiang 2013) Parker winds typically operate when the sound speed, $c_s$, at the base of the atmosphere is $\sim$5 times lower than the escape speed.  For typical heavy molecules that will form grains, the sound speed is $c_s \simeq 1\times \sqrt{T/2000\, {\rm K}}$ km s$^{-1}$, while the escape speed from an Earth-size planet is $v_{\rm esc} \simeq \sqrt{2GM_p/R_p} \simeq 10 $ km s$^{-1}$.  This fiducial ratio of $v_{\rm esc}/c_s$ is thus $\sim$10.  In order to reduce this by a factor of $\gtrsim$ 2 to be more squarely in the Parker-wind regime means decreasing the planet mass by approximately a factor of $\gtrsim$ 8 since, at constant density, $v_{\rm esc}$ scales as $M_p^{1/3}$.  But, roughly speaking, decreasing an Earth-size rocky planet's mass and radius by factors of $\gtrsim$ 8 and $\gtrsim$ 2, respectively, puts it in the Mars to Mercury mass and size range.  In fact, Perez-Becker \& Chiang (2013) find the optimal mass-loss solutions (at least for KIC 1255b) to involve planet masses in the range of $0.01-0.02\,M_\oplus$, corresponding to radii of $\sim$$0.22-0.28 \,R_\oplus$, and $v_{\rm esc}/c_s $ as low as $\sim$2.  Thus, if KOI-2700b is indeed a rocky planet that is in the final catastrophic phase of disintegration via dust emission (see Perez-Becker \& Chiang 2013), it must be quite small and low in mass.  

In this regard, Kepler-10b and Kepler-78b, with well-studied properties (Batalha et al.~2011; Sanchis-Ojeda et al.~2013a; Howard et al.~2013; Pepe et al.~2013), are both short-period ($P_{\rm orb} = 0.83$ and 0.36 days, respectively) rocky planets that do {\em not} exhibit any evidence for dust emission in their transits.  However, their surface gravities are $\sim$2.3 $g_\oplus$ and $\sim$1 $g_\oplus$, respectively, and no Parker-wind is expected to develop in their atmospheres.  Hence, this is perfectly consistent with them showing no evidence for dusty emissions.

In keeping with this scenario, we would {\em not} expect any ``normal-looking'' transit, with an inferred rocky planet radius 
that is $\gtrsim$ 1/2 $R_\oplus$, to exhibit any hint of a dusty tail.  Conversely, if one detects an asymmetric transit, due to a dusty tail, then it will be very difficult to detect the hard body of the planet within the transit unless the host star is quite bright and the photon statistics are extraordinarily high because, by necessity, the planet must be quite small (i.e., $\lesssim 0.3 \,R_\oplus$).  Moreover, given the complications of the transit shape produced by a dust tail, it may be nearly impossible to detect the hard-body of the planet even with the best of statistics.  

Finally, we note that in this work we have not considered models for the asymmetric transit profile other than that caused by a dusty comet-like tail.  This is due to the similarities of KOI-2700b with KIC 1255b, and the fact that no satisfactory alternative models have thus far been proposed to explain the asymmetric tail in that system.  In this regard, we note that the 0.44-day planet PTFO-8-8695b (van Eyken et al.~2012; Barnes et al.~2013) also exhibits some asymmetric transits, but with a slow ingress and rapid egress, the opposite of what is seen in KIC 1255b and KOI-2700b.  Barnes et al.~(2013) attribute the asymmetric transit profiles to a planet whose orbital plane is misaligned with the rapidly rotating host star, thereby leading to transits that are angled across the severely gravity darkened equatorial regions of the star.  That scenario, if correct, cannot apply to either KIC 1255b or to KOI-2700b because (i) the rotation periods of those host stars are 20-40 times longer, and (ii) the transits, including the long egress tails, are substantially longer in duration than the passage of a hard-body planet would be across the disk of either host star.  Nonetheless, this object (PTFO-8-8695b)  provides at least one example of an asymmetric transit that can be explained without invoking a dust tail.

\acknowledgements We thank Eugene Chiang and Josh Winn for helpful discussions. R.S.O. acknowledges NASA support through the {\em Kepler} Participating Scientist Program.  This research has made use of data collected by the {\em Kepler} mission, which is funded by the NASA Science Mission directorate.  Some of the analysis made use of {\tt PyKE} (Still \& Barclay 2012), a software package for the reduction and analysis of {\em Kepler} data.  This open source software project is developed and distributed by the NASA {\em Kepler} Guest Observer Office.  The J-band image of the KOI-2700 field was obtained with the United Kingdom Infrared Telescope (UKIRT) which is operated by the Joint Astronomy Centre on behalf of the Science and Technology Facilities Council of the U.K.  Some of the data presented in this paper were obtained from the Mikulski Archive for Space Telescopes (MAST).  STScI is operated by the Association of Universities for Research in Astronomy, Inc., under NASA contract NAS5-26555.  Support for MAST for non-HST data is provided by the NASA Office of Space Science via grant NNX13AC07G and by other grants and contracts.

\appendix

\section{Approximate Exponential Decay of Cross Section With Time}
\label{app:appA}

\subsection{Equation of Dust Heating and Cooling}
\label{sec:heat_cool}

Following Kimura et al.~(2002), we write down the equation for determining the equilibrium temperature of dust grains for the assumptions that they (i) are simple spheres, (ii) remain at a fixed distance from the parent star, and (iii) come into thermal equilibrium on a timescale short compared to their evaporative lifetime.
\begin{eqnarray}
\frac{\pi R_s^2}{d^2}\int_0^\infty \sigma_a(a,\lambda)B(\lambda,T_s)d\lambda= \int_0^\infty 4 \pi^2 a^2 \epsilon(a,\lambda)B(\lambda,T_d)d\lambda+\mathcal{L} \frac{dm_d}{dt} 
\label{eqn:heat1}
\end{eqnarray}
where the left-hand side represents the heating of a dust grain at a distance $d$ from the host star, and the two terms on the right hand side are the radiative cooling of the dust grain and cooling via sublimation, respectively.  The subscripts $s$ and $d$ stand for the host star, and the dust, respectively.  The other variables are the absorption cross section, $\sigma_a$; the emissivity of the dust grain, $\epsilon$; the latent heat of sublimation, $\mathcal{L}$; and the mass loss rate of the grain, $dm_d/dt$, which we take to be a positive quantity.  By Kirchoff's law, $\epsilon(a,\lambda) = \sigma_a(a,\lambda)/\pi a^2$.

The mass-loss term is
\begin{eqnarray}
\frac{dm_d}{dt} = 4 \pi a^2 \sqrt{\frac{\mu}{2 \pi k T_d}}P_0e^{-\mu \mathcal{L}/k_BT_d}
\label{eqn:mdot}
\end{eqnarray}
where $\mu$ is the mean molecular weight and $P_0$ is the vapor pressure as $T \rightarrow \infty$.   

If one considers olivines and pyroxenes, and then plots the left-hand side of Eqn.~(\ref{eqn:heat1}) vs.~the two terms on the right-hand side, individually, it becomes clear that the first term on the r.h.s.~equals the l.h.s.~well before the second term on the r.h.s.  This indicates that the equilibrium temperature is set by the following much simpler equation:
\begin{eqnarray}
\frac{R_s^2}{d^2}\int_0^\infty \sigma_a(a,\lambda)B(\lambda,T_s)d\lambda \simeq \int_0^\infty 4 \sigma_a(a,\lambda) B(\lambda,T_d)d\lambda
\label{eqn:heat2}
\end{eqnarray}
Based on the results of Fig.~\ref{fig:sigma} (and similar calculations for other small, but constant, values of $k$; not shown) we find that, to a decent approximation:
\begin{eqnarray}
 \sigma_a(a,\lambda) \simeq {\rm constant} \times \pi a^2 \left(\frac{2 \pi a}{\lambda}\right)
\label{eqn:sigma_a}
\end{eqnarray}
for small particle sizes.  The ``constant" in this expression depends linearly on the choice of $k$, the complex index of refraction (at least for small values of $k$).  Equation \ref{eqn:heat2} then reduces to:
\begin{eqnarray}
\left( \frac{R_s}{2d} \right)^2 \int_0^\infty \frac{1}{\lambda} B(\lambda,T_s)d\lambda \simeq \int_0^\infty \frac{1}{\lambda} B(\lambda,T_d)d\lambda
\label{eqn:heat3}
\end{eqnarray}
After multiplying and dividing by $T$, the integrals on both sides of the equation become dimensionless and cancel, leaving 
\begin{eqnarray}
T_d \simeq \left( \frac{R_s}{2d} \right)^{2/5} T_s 
\label{eqn:Tdust}
\end{eqnarray}
an approximate result that is independent of particle size and composition.  If we apply this to KOI-2700, we find $T_d \simeq 1650$ K.

\subsection{Cross Section Decaying With Time}
\label{sec:decay}

To the extent that the grain temperatures are independent of size, the mass-loss rate, $\dot m_d$, given by Eqn.~(\ref{eqn:mdot}) implies that $\dot m_d$ is simply proportional to the surface area, $a^2$, for all grains of a given composition (i.e., sharing common values of $\mu$ and $\mathcal{L}$).  But, for a fixed bulk density $\rho$
\begin{eqnarray}
\frac{da}{dt} \simeq \frac{\dot m_d}{4 \pi a^2 \rho} \simeq \mathcal{R}(\mu,\mathcal{L}, d, T_s, R_s)
\label{eqn:adot}
\end{eqnarray}
where $\mathcal{R}$ is a `constant' that is understood to depend on the chemical composition of the dust and the properties of, and distance from, the host star.

\begin{figure}
\begin{center}
\includegraphics[width=0.60 \columnwidth]{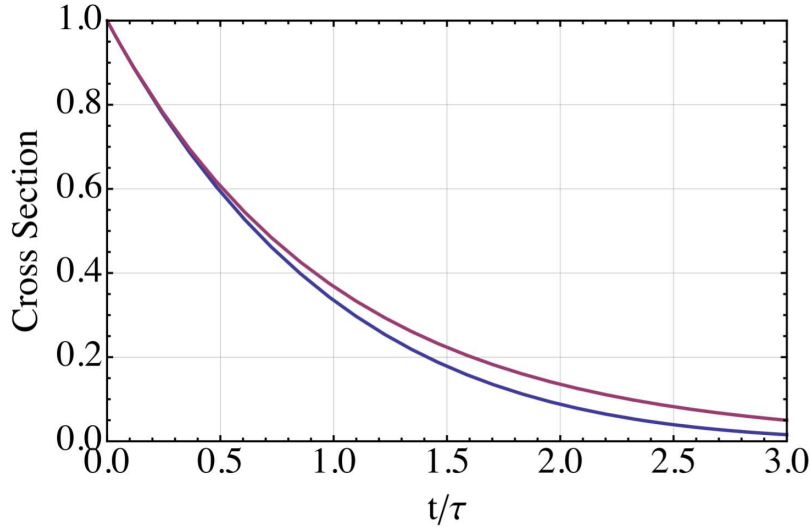} 
\caption{Comparison of the functions $(1-t/6\tau)^6$ with $e^{-t/\tau}$. The top (red) curve is the exponential.}
\label{fig:expon}
\end{center}
\end{figure}

For small particle sizes, i.e., $2 \pi a/\lambda \lesssim 2$, the extinction cross section (relevant for producing the transit profile) goes like the Rayleigh limit with:
\begin{eqnarray}
\sigma_{\rm ext} ~\propto~\frac{a^6}{\lambda^4}
\label{eqn:extinct}
\end{eqnarray}
(see Fig.~\ref{fig:sigma}).  At wavelengths of $\sim$0.5 to 1 $\mu$m, this approximation requires particle sizes $\lesssim 0.25 \, \mu {\rm m}$.  Thus, the rate of change of the extinction cross section with time, as small particles shrink due to sublimation, is given by
\begin{eqnarray}
\frac{ d\sigma_{\rm ext}}{dt} ~\propto~a^5 \dot a ~\propto~- a^5 \mathcal{R} ~\propto ~-\sigma_{\rm ext}^{5/6} ~\equiv~-\frac{\sigma_{\rm ext}^{5/6} \sigma_0^{1/6}}{\tau} 
\label{eqn:sigma_dot}
\end{eqnarray}
where the last expression on the right makes use of two definitions: $\sigma_0$ is the cross section before the grain starts to sublimate, and $\tau$ is a characteristic decay time which is proportional to size, $a$.  Thus, $\sigma_s$ has the following time dependence:
\begin{eqnarray}
\sigma_{\rm ext} = \sigma_0 \left(1-\frac{t}{6 \tau} \right)^6
\label{eqn:sigma}
\end{eqnarray}
However, since in the limit of $n \rightarrow \infty$, $(1-t/n \tau)^n \rightarrow e^{-t/\tau}$, we find that an integer $n = 6$ is sufficiently large to make the exponential solution a decent approximation to the time dependence of the cross section on time.  A plot comparing the exponential solution with the $n = 6$ solution is shown in Fig.~\ref{fig:expon}.
In this expression, the cross section depends sensitively on size, $a$, but the decay timescale, $\tau$, is just linearly dependent on $a$.  Thus, if the distribution of grain sizes in the dust tail is limited at the high end by the largest particles that are readily ejected from the planet, and from below by those that decay away quickly, there may be a sufficiently narrow range of particle sizes so that a simple exponential function is a relatively reasonable approximation to the decay of the effective cross section of the particles in the dust tail.

\section{Estimates of Forces and Heating by Stellar Winds}
\label{app:winds}

\subsection{Ram Pressure Forces on Dust}
\label{sec:ram}

For neutral dust particles that are not affected by the magnetic fields entrained in the stellar wind, we consider only the direct ram pressure of the protons and helium atoms colliding with the dust grains.  The radial component of the force is given by
\begin{eqnarray}
F_w = \pi a^2 \,\mu \,n_w \,v_w^2
\label{eqn:ram}
\end{eqnarray}
where the subscript $w$ indicates the stellar wind, and $\mu$, $n_w$, and $v_w$ are the mean weight, number density, and velocity of the stellar wind particles, respectively.  The ratio of this force to gravity is given by
\begin{eqnarray}
\beta_w \equiv \frac{F_w}{F_{\rm grav}} = \frac{3 \mu v_w^2 n_w d^2}{4 GM_s a \rho} \simeq 0.001  \left(\frac{v_w}{200\,{\rm km}\,{\rm s}^{-1}}\right)^2\left(\frac{n_w}{10^6 \, {\rm cm}^{-3}}\right)\left(\frac{d}{5\,M_\odot}\right)^2\left(\frac{M_\odot}{M_s}\right)\left(\frac{\mu{\rm m}}{a}\right)\left(\frac{4 \, {\rm g\,cm}^{-3}}{\rho}\right)
\label{eqn:ram}
\end{eqnarray}
where $\rho$ is the bulk density of the grain material, and we have normalized to plausible parameter values for the stellar wind from KIC 8639908 in the vicinity of KOI-2700b; though these are obviously highly uncertain (see, e.g., Cohen et al.~2011; Bisikalo et al.~2013).  We see that the contribution to $\beta$ from the stellar wind ram pressure, $\beta_w$, will be one to two orders of magnitude smaller than due to radiation pressure (see Sect.~\ref{sec:beta}) unless the stellar wind density is extremely and unexpectedly high without the value of $v_w$ also being smaller than 200 km s$^{-1}$.  We note that the implied mass loss rate with these normalization parameters is $2.3 \times 10^{13}$ g s$^{-1}$, or about an order of magnitude {\em higher} than for the Sun.  This is plausible for a mid-K star with an angular frequency that is 3 times higher than that of the Sun.

\subsection{Heating of Grains by the Stellar Wind}
\label{sec:heat}

Under the same set of assumptions as in Sect.~\ref{sec:ram} the collisional heating of the dust grains from the stellar wind will be
\begin{eqnarray}
\dot Q_w = \frac{1}{2} \pi a^2 \,\mu \,n_w \,v_w^3  \simeq 2 \times 10^{-4}  \left(\frac{a}{\mu{\rm m}}\right)^2  \left(\frac{n_w}{10^6 \, {\rm cm}^{-3}}\right)  \left(\frac{v_w}{200\,{\rm km}\,{\rm s}^{-1}}\right)^3  ~{\rm ergs~s}^{-1}
\label{eqn:heat}
\end{eqnarray}
where we have normalized to the same estimated parameters as in the previous section.
In the absence of heating by radiation, the equilibrium temperature of the dust grains due to collisional heating by the stellar wind would be 
\begin{eqnarray}
T_{d,w} \simeq 330  \left(\frac{\epsilon}{0.01}\right)^{-1/4}  \left(\frac{n_w}{10^6 \, {\rm cm}^{-3}}\right)^{1/4}  \left(\frac{v_w}{200\,{\rm km}\,{\rm s}^{-1}}\right)^{3/4}    ~{\rm K}
\label{eqn:temp}
\end{eqnarray}
where we have simply taken the emissivity to be grey with an overall value of 10\%, independent of particle size or wavelength.  Thus, the stellar wind heating would be insignificant compared with radiative heating except in the case of implausibly low emissivities or high stellar wind densities.

\end{document}